\documentclass[preprint,12pt]{elsarticle}

\usepackage{graphicx}
\usepackage{ifpdf}
\usepackage{pdfpages}

\usepackage{amssymb}
\usepackage{amsmath}

\usepackage{color}

\usepackage{hyperref}

\begin{document}

\begin{frontmatter}



\title{Duplex formation and secondary structure of $\gamma$-PNA observed by NMR and CD}


\author[UdS]{J.M.P. Vi\'eville}
\author[Geneva]{S. Barluenga}
\author[Geneva]{N. Winssinger}
\author[IGBMC]{M-A. Delsuc\corref{cor1}}
\ead{delsuc@igbmc.fr}
\address[UdS]{Strasbourg University, Plateforme d'Analyse Chimique de Strasbourg Illkirch, 74 route du Rhin 67401 ILLKIRCH, FRANCE}
\address[Geneva]{Department of Organic Chemistry, University of Geneva, Geneva CH1211, SWITZERLAND}
\address[IGBMC]{IGBMC, CNRS UMR 7104, 1 rue Laurent Fries BP10142, 67404 ILLKIRCH FRANCE}

\begin{abstract}

Peptide Nucleic Acids (PNA) are non-natural oligonucleotides mimics, wherein the phosphoribose backbone has been replaced by a peptidic moiety (N-(2-aminoethyl)glycine).
This peptidic backbone lends itself to substitution and the $\gamma$-position has proven to yield oligomers with enhanced hybridization properties.
In this study, we use Nuclear Magnetic Resonance (NMR) and Circular Dichroism (CD) to explore the properties of the supramolecular duplexes formed by these species.
We show that standard Watson-Crick base pair as well as non-standard ones are formed in solution.
The duplexes thus formed present marked melting transition temperatures substantially higher than their nucleic acid homologs.
Moreover, the presence of a chiral group on the $\gamma$-peptidic backbone increases further this transition temperature, leading to very stable duplexes.

PNA duplexes with a chiral backbone present a marked chiral secondary structure, observed by CD, and showing a common folding pattern for all studied structures.
Nevertheless small differences are observed depending on the details of the nucleobase sequence.

\end{abstract}


\end{frontmatter}


\section*{Introduction}
\label{intro}

Peptide Nucleic Acid (PNA) are synthetic oligonucleotides first reported by Nielsen\cite{Nielsen:1991tu}
 with a backbone that recapitulate DNA's inter-nucleobase distances.
PNAs were developed as DNA mimics in order to recognize DNA double helix\cite{Anonymous:NrVprpQz} and to form hybrid duplexes with DNA\cite{Egholm:1993eo} or homoduplexes\cite{10.1038_368561a0}.
PNAs duplexes have high thermal stability,
are not sensitive to nucleases or protases, and are metabolically stable.
However, as polyamide backbone has no charge, no Coulomb repulsion is reported and PNA aggregates may exists in solution, and their solubility in water is limited\cite{Uhlmann:1998uz}.
PNAs have attracted interest in molecular biology and numerous applications have been reported,  we can cite beacons for duplex DNA\cite{Kuhn:2001ep},
and PNA as a diagnostic tool\cite{doi:10.1586/14737159.3.5.649},
a genomic tool\cite{doi:10.1586/14737159.1.3.343},
or a supramolecular barcodes\cite{Pianowski2008,10.1021/acs.accounts.5b00109}.

These applications were leveraged on the unique properties of PNA hybridizing to DNA or RNA.
The enhanced hybridization properties of $\gamma$-modified PNA have been recently harnessed for gene editing applications\cite{10.2174/1566523214666140825154158}.
PNA have also been used to tag small or macromolecules and program their assemblies based on hybridization\cite{Pianowski2008} \cite{10.1021/acs.accounts.5b00109}.
PNA homoduplexes have also been used in programmed assemblies as recently illustrated for the programmed pairing of PNA-tagged protein fragments\cite{Kazane:2013jx}.
While interactions of PNA with DNA or RNA have been extensively studied, homoduplex formation of PNA are not well characterized, particularly for modified PNAs.

PNA-DNA hybridization by Watson-Crick base pairing are well studied\cite{Egholm:1993eo}.
They reveal a helix formed by the $\gamma$-carbon and the nitrogen of the tertiary amide of the PNA backbone\cite{Avitabile:2012ed}, stabilized by the sequential base stacking\cite{Yeh:2011gs}.
PNA neutral backbone bring thermal stability in PNA/DNA duplex compared to DNA/DNA duplexes, and present a better specificity\cite{Demidov:2004hj}.
Because of the absence of charge on the PNA backbone, the PNA/DNA hybrids are not dependent on
the ionic strength of the solvent in contrast to DNA homoduplexes.
Because of a flexible backbone, Watson-Crick, Hoogsteen, reverse Hoogsteen, or Wobble interactions can be formed
by PNA/DNA duplexes\cite{Faccini:2008jk}. 

Unlike DNA which has a helical conformation due to its chiral centers, 
standard PNAs, with no chiral carbon, have no defined structural conformation in solution.
Adding a chiral center on the PNA backbone forces left or right helix\cite{DragulescuAndrasi:1970hq}, and increases PNA/DNA hybridization.
Adding a Lysine improves water solubility\cite{Yeh:2011gs}, however the Serine is less disruptive to hybridization and --OH groups can form hydrogen bonds with water in order to increase solubility.
With the help of chiral centers, PNA homoduplexes should also be observed, and eventually present secondary helical structures.
In this study, we explore the biophysical properties of PNAs built with a chiral center.
The chiral center chosen here, consists in a L-Serine substitution of the $\gamma$-position on the PNA backbone\cite{He:2012dy}, see insert in figure \ref{pna_molec}.

In order to determine homo-hybridization of PNAs through a structural pre-organization, a series of $\gamma$-PNAs were analyzed by Circular Dichroism (CD) and Liquid State Nuclear Magnetic Resonance (NMR).
NMR allows to spotlight hydrogen bounds formed by nucleic acid base-pairing.
Working in neutral conditions, specific NMR experiments bring information about the imino hydrogens engaged in the Watson-Crick, Hoogsteen, or Wobble bounds.
Melting transitions ($T_m$) of the highlighted PNA homoduplexes were determined by CD and absorbance experiments.
Circular Dichroism is finally used to assess the secondary structure displayed by the homoduplexes.
All the experimental results show unambiguously that PNAs duplexes are formed, and that the chiral centers improve their stability, and drive the duplexes into a chiral secondary structure, probably organized as a left-handed helix.

\section*{Experimental} 
\label{sec:experimental}

\subsection*{Materials} 
\label{sub:materials}

\begin{figure}
\includegraphics[width=1\textwidth]{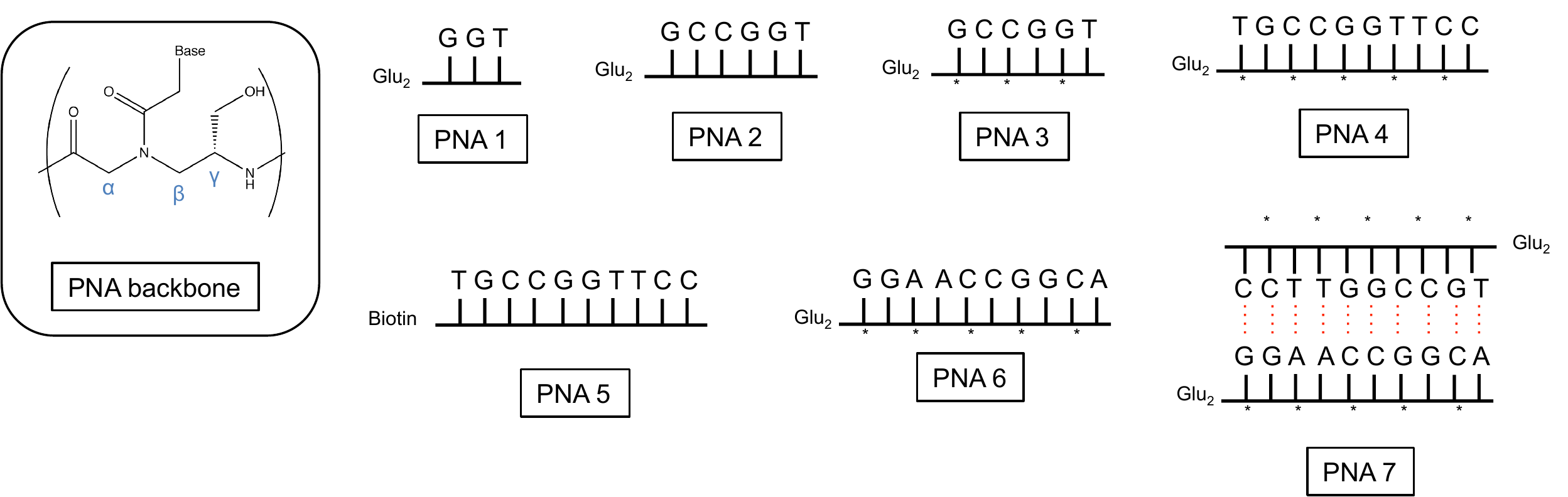}
\caption{Schematic view of the 6 PNAs analyzed. 
Insert shows PNA backbone and the $\gamma$ position of L-Serine.
Stars represent chiral centers  on the PNA backbone.
PNA \underline{2} and \underline{3} have the same nucleobases sequence but different backbones: \underline{3} is a  $\gamma$-PNA with 3 chiral centers.
PNA \underline{4} and \underline{5} have the same nucleobase sequence.
PNA \underline{6} is the complementary PNA of \underline{4},
and PNA \underline{7} is the \underline{4}--\underline{6} duplex.
Note that only PNA \underline{5} has a Biotin and not a Glutamine at the end of the backbone chain. 
PNA \underline{1}, \underline{2} and \underline{5} have no chiral centers.
}
\label{pna_molec}
\end{figure}

PNAs were synthesized as previously reported and purified by HPLC\cite{Chouikhi2012}. 
Six PNAs (see figure \ref{pna_molec} with a schematic view) were chosen according to their nucleobases and the presence or absence of $\gamma$ modification on the backbone.
PNA \underline{1} is a small one, with three nucleobases \texttt{GGT} and has a standard backbone.
PNA \underline{2} and \underline{3} have the same six nucleobases: \texttt{GCCGGT}, differing by the presence of the $\gamma$ L-Serine on PNA 3 schematically represented by a star.
PNA \underline{4} and \underline{5} are also similar PNAs, their backbone contain ten nucleobases \texttt{TGCCGGTTCC}, and differ only by the presence of the  $\gamma$ L-Serine on PNA \underline{4}.
PNA \underline{6} is the complementary strand of PNA \underline{4} and \underline{5}, and presents also chiral centers.

First, PNAs were dissolved in Milli-Q water, and diluted into
 phosphate buffer  (pH=6.8).
10\% of deuterated water was added and
10\% of deuterated DMSO was also added to PNA \underline{5} solution to insure solubility.
Before study, all solutions were exposed to several annealing cycles by using a solid bath, and raising the temperature from room temperature to 95\,$^\circ$C and slowly cooled back to room temperature.

\subsection*{NMR spectroscopy} 
\label{sub:nmr_spectroscopy}

PNAs were analyzed in 3\,mm tubes on a 700MHz Bruker spectrometer equipped with a Z-grad triple resonance cryoprobe.
PNAs were prepared as above, with 10\% D$_2$O added for lock.
\textit{Watergate} or \textit{Excitation Sculpting} water suppression experiments were not used because of the concern that the exchangeable imino protons would be suppressed too.
The 1D suppression experiments used here is the \textit{Jump \& Return} (\emph{JR}) excitation sequence\cite{Plateau:1982vj}, 
followed by a \emph{W5} water removal sequence\cite{W5}.
The \textit{JR} sequence suppresses the water signal by not exciting the water spin, in consequence the water magnetization is unperturbed and exchangeable protons like the imino protons appear fully in the 1D $^1$H NMR spectrum.
The \emph{W5} sequence cleans the spectrum from residual water signal, with no perturbation on the other signals. 


NMR experiments were run at a various temperatures, between 288\,K and 318\,K,
as noted in the figure captions.
In order to investigate the structural conformation and to assign spectra, \textit{JR}-NOESY, HSQC, TOSCY and DOSY spectra were recorded on each compound.
Spectra not presented in the text can be found in the Supplementary Materials.
PNA concentrations range from 0.88\,mM to 1.71\,mM depending on the experiments and the studied PNA.


\subsection*{Optical Measure} 
\label{sub:opti_meas}

\paragraph{Absorbance}\

PNA \underline{3}, \underline{4} and \underline{5} were analyzed at 60\,$\mu$M (PNA \underline{3}) and 50\,$\mu$M (PNA \underline{4} - \underline{5}) by Absorbance on a Jasco J-815 spectropolarimeter in 1\,mm path length quartz UV cells.

$T_m$ determinations were performed by absorbance measurement as a function of the temperature. 
From an initial absorbance spectrum, five wave-lengths were chosen and followed during the temperature variation.
Absorbance curves were recorded at several positions between 200\,nm and 350\,nm.
 corresponding to the highest absorbance, or the lowest absorbance (blank signal).
Absorbance was monitored during heating, the temperature varying from 308\,K to 368\,K with a 1\,K/mn variation rate, and a cooling experiment performed at the same rate.
The $T_m$ were determined by a fitting procedure of the various absorbance and CD curves obtained, using a python program (see Supplementary Materials).
The program is freely available at \url{https://github.com/delsuc/Melting-curve-analysis}

\paragraph{Circular Dichroism}\

PNA \underline{3}, \underline{4}, \underline{6}, and \underline{7}  were analyzed at 60\,$\mu$M (PNA \underline{3}) and 50\,$\mu$M (PNA \underline{4 - 6 - 7}) by Circular Dichroism on the same Jasco equipment.

Samples were heated from 308\,K to 363\,K, and cooled down from 363\,K to 308\,K with the same temperature slope.
CD spectra were first measured at 308\,K before the temperature variation, then at 363\,K, and finally again at the low 308\,K after the cooling procedure.

Melting transitions were compared to prediction obtained for RNA presenting the same sequence, 
using the DINAMelt Web Server (Di-Nucleic Acid hybridization and Melting prediction).
\cite{Markham01072005}
This tool predicts nucleic acid pairing and melting transitions for DNA or RNA strands.
It is found at \url{http://mfold.rna.albany.edu/?q=dinamelt}.
Schematic view of PNA homo-hybridization proposed by the program are shown in Supplementary Materials (Figure S14).



\section*{Results and Discussion} 
\label{sec:results}
\subsection*{NMR solution analysis of PNAs} 
\label{sub:nmr}

\begin{figure}
\centering
\includegraphics[width=1\textwidth]{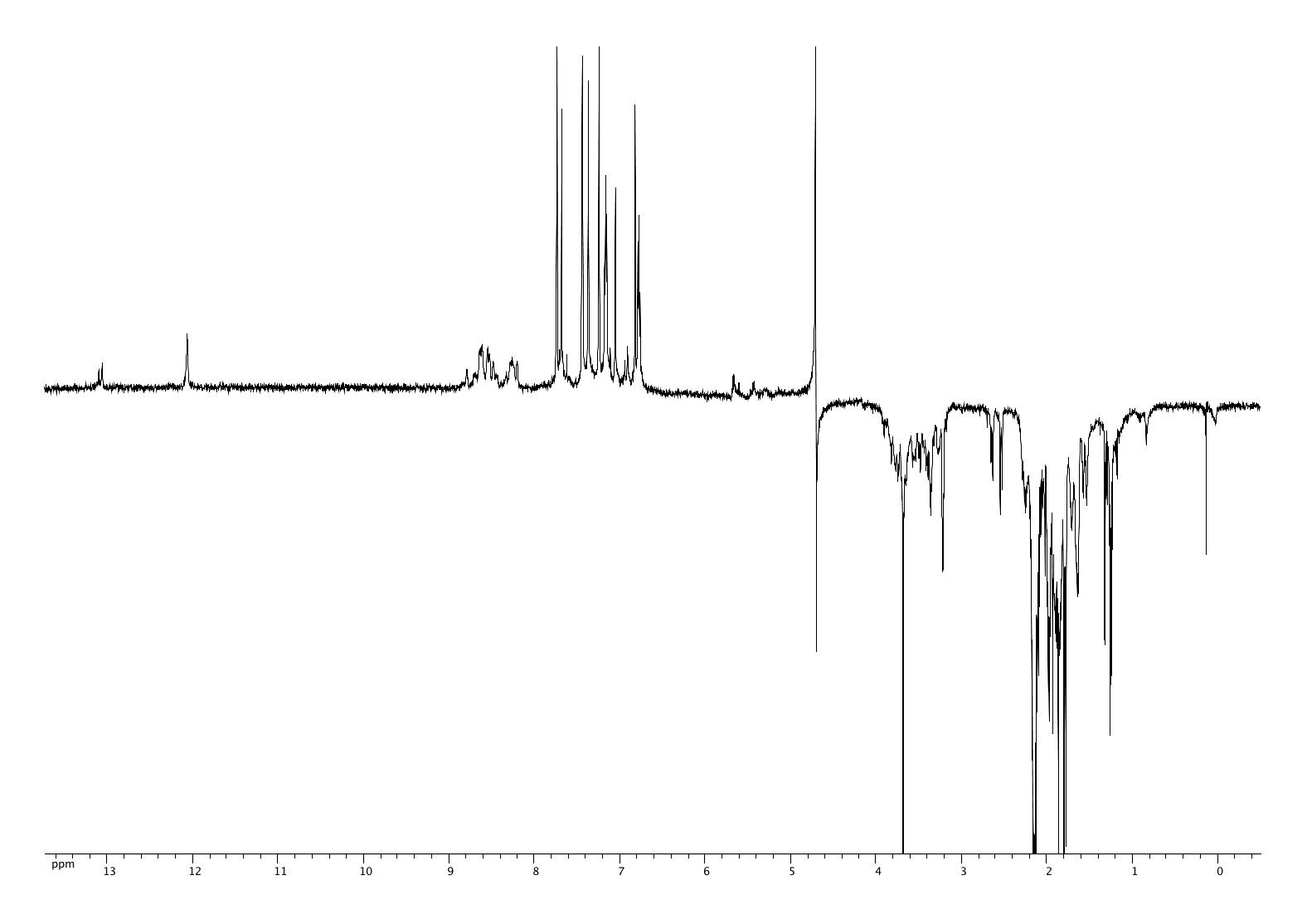}
\caption{1D $^1$H \textit{JR} NMR spectra of PNA \underline{3}.
NS=196, [\underline{3}]=1.1mM, [Phosphate]=25mM, T=298K, 700MHz.
Signals at 12.9--13\,ppm correspond to imino protons involved in G-C base pairing, and signals at 12\,ppm correspond to imino proton from T-G base pairing.}
\label{3_jr}
\end{figure}

The various PNA constructs were first studied by NMR.
The NMR spectrum of PNA \underline{3} is shown in figure \ref{3_jr}.
Because of the \textit{JR} sequence used, the spectrum presents two domains, a positive one and a negative one, on each side of the location of the water signal.
The use of the \emph{JR} sequence allows the observation of the exchangeable protons which would be lost using a standard 1D $^1$H excitation.

By homology with the NMR signals of DNA duplexes, the 13\,ppm resonance signals are assigned to the imino proton of G-C Watson-Crick base pairs.
Signals at 12\,ppm are assigned to the imino proton of a Wobble base pair between a G nucleobase and a T nucleobase.
From the NMR signals alone, it is not possible to determine whether the spectrum corresponds to a duplex formation or to a folding of the PNA on itself in a hair-pin conformation.
The DOSY experiment performed on this molecule has been inconclusive (see Supplementary Data S3).
However, the presence of two \mbox{G-C} base pairing seems incompatible with a hair-pin geometry, and probably indicates that PNA \underline{3} is in a homoduplex form in solution.

The comparison of the integrals of the imino and amino protons is difficult because the excitation sequence used here has a non-uniform excitation profile over the spectrum, however it can be estimated from the respective integrals that there is a default of imino signals in the spectrum.
As imino protons are specific to stable base-pairs, this is indicative that some amount of the PNA material is still in a single strand form in solution.



\begin{figure}
\centering
\includegraphics[width=0.95\textwidth]{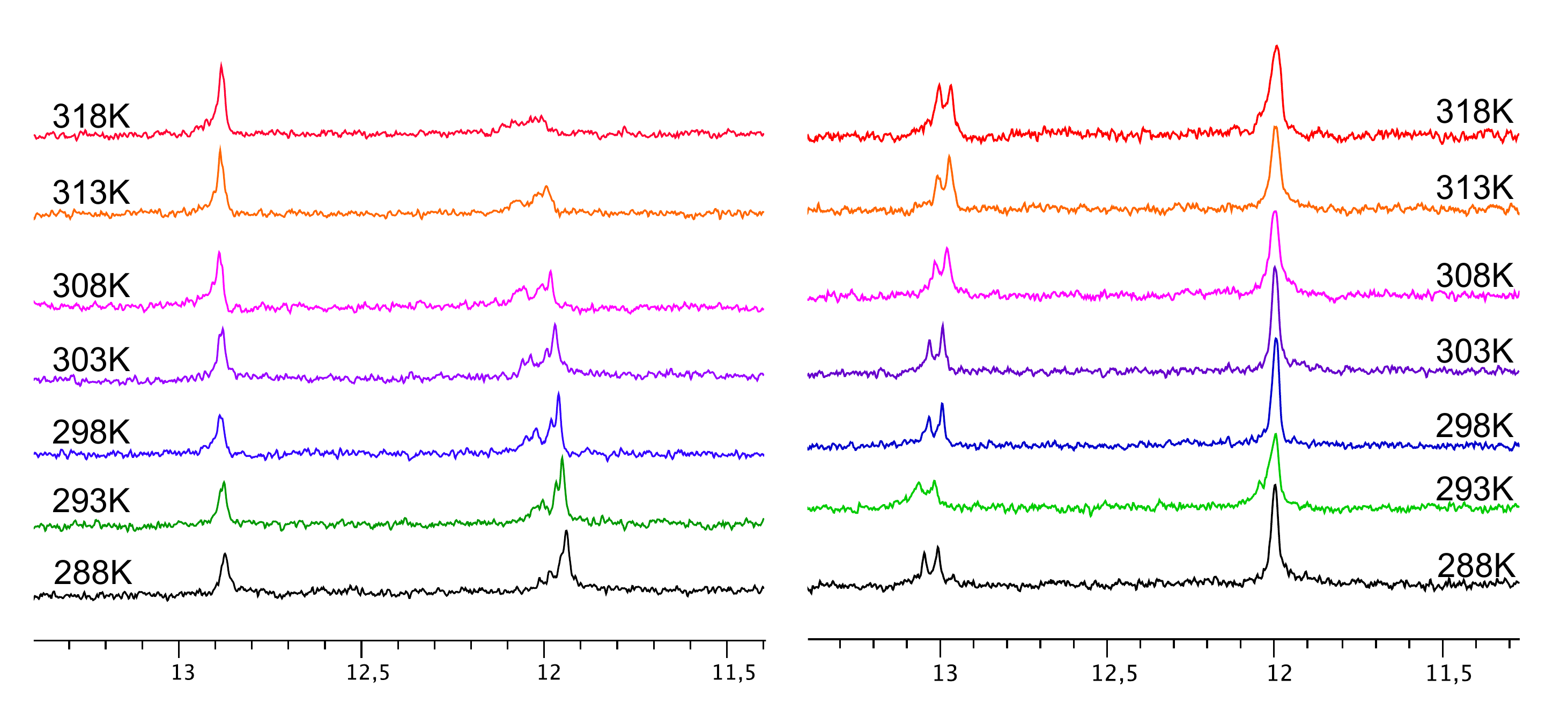}
\caption{1D proton \textit{JR} NMR spectra of $\gamma$-PNA \underline{2} (left) and PNA \underline{3} (right) at various temperatures, zoomed in the 11.4--13.4\,ppm area.
[\underline{2}]=0.88mM, [\underline{3}]=1.1mM, [Phosphate]=25mM, 
Signals at 12.9--13\,ppm correspond to G-C base pairing, and signals at 12\,ppm correspond to T-G base pairing.
}
\label{1Djr}
\end{figure}

NMR spectra of PNA \underline{2} and \underline{3} for a temperature variation are shown in figure \ref{1Djr}.
The figure presents a zoom of the imino protons area, and spectra are recorded for temperatures ranging from 288\,K to 318\,K.
For PNA \underline{3}, NMR spectra at various temperatures do not present any differences,
in contrast to PNA \underline{2} for which differences are seen for the T-G imino proton signal with a decrease of the signal at high temperature.
For PNA \underline{2} the T-G signal disappears completely at 318\,K, unlike the G-C signals, which are still present.
This indicates a stronger binding for the G-C base pair compared to the T-G base pair.
As PNA \underline{2} and PNA \underline{3} contain the same nucleobase sequence, this indicates that the presence of the L-Serine in the $\gamma$ position strengthen the Wobble T-G pair and increases the overall stability of the PNA homoduplex.
The same temperature variation experiment has been made for PNA \underline{4} and \underline{5}, (see figures S12 and S13 Supplementary Material) and lead to the same conclusion.

The NMR instrument cannot go higher in temperature due to the limit on the equipment, 
and the melting transition is not visible in this experiment.
Optical approach were thus utilized to determine precisely the melting transition of the different PNAs.


\subsection*{Secondary structural analysis} 
\label{sub:secondary_structural_analysis}

The melting-transition temperature $T_m$ of the various $\gamma$-PNA studied here have been determined. Melting temperature were determined by temperature variation, monitoring both the absorbance and the ellipticity at several different wavelengths (see Supplementary Materials S17--S35).
The experimental values are presented in the table \ref{table:tm}.
They are compared to theoretical $T_m$ values that nucleic acid strands with the same sequence would present, as computed using the DINAMelt Web Server\cite{Markham01072005}.
Values computed for RNA were considered, because the enhanced flexibility of the RNA backbone better matches the flexibility of the $\gamma$-PNA structure.

Except for PNA \underline{1} which do not present any duplex formation, the different PNAs studied here present strong base pairing, with rather high melting temperatures.
The addition of a chiral center, in the form of a L-Serine substitution in the $\gamma$ position on the PNA backbone has a stabilizing effect.
This substitution has no net effect in the case of PNAs \underline{2} and \underline{3} but raises by 14 degrees the longer PNAs \underline{5} and \underline{4}.
The duplex \underline{7}, with 10 potential base-pairs and a chiral backbone for both strands, presents an high stability, with a transition temperature well above 100\,$^\circ$C.
Such an extreme stability implies that we cannot be sure of the complete pairing of the strands in this sample, as the annealing cycles, performed in water, cannot completely disrupt the duplex in order to allow a complete sampling of all the possible conformations.

For all the PNAs tested here, the experimental melting temperatures are quite higher than the computed values of their RNA counterpart.
This is probably due to the absence of the strong Coulombic repulsion observed in RNA because of the charged backbone.
It should be noted that the differences between the experimental and computed values present important differences for the considered PNAs.
In the case of PNAs \underline{2} and \underline{3}, with four Watson-Crick G-C and 2 Wooble T-G base pairs, this difference is 30 to 35\,$^\circ$C.
PNAs \underline{4} and \underline{5} have the same potential base-paring, located on a longer strand.
They present differences on the same order of magnitude, except may be for PNA \underline{5} which lacks chiral centers, and for which the entropic effect of the long non paired strand may have some destabilization effect.
PNA \underline{6} on the other hand is puzzling.
While it is expected to form only 4 Watson-Crick G-C base pairs, it presents a melting transition equivalent to PNA \underline{4}, 58\,$^\circ$C above its theoretical RNA counterpart.
This cannot be explained without the formation of additional non-standard base pair, such as the C$-\mspace{-6mu}{\bullet}\mspace{-6mu}-$A or A$-\mspace{-6mu}{\bullet}\mspace{-6mu}-$A base pairs, as observed in some RNA secondary structures\cite{Leontis:2002ts}.
These non-standard pairs are not isosteric to the standard Watson-Crick ones, but their formation might be possible here thanks to the greater flexibility on the PNA backbone.


\begin{table}
\begin{center}
\begin{tabular}[h]{c|ccc}
\hline
& Exp $T_m$ & computed $T_m$ & Diff \\
\hline \hline
PNA \underline{2} & 341 K $\pm$ 2.0 & 308.8 K & 32.2 \\ 
PNA \underline{3} & 341 K $\pm$ 1.8 & 308.8 K & 32.2 \\ 
PNA \underline{4} & 359 K $\pm$ 3.4 & 323.8 K & 35.2 \\ 
PNA \underline{5} & 345 K $\pm$ 1.9 & 323.8 K & 21.2 \\ 
PNA \underline{6} & 357 K $\pm$ 2.8 & 298.4 K & 58.6 \\ 
PNA \underline{7} & $\geq$ 383 K   & 342.0 K & $\geq$ 40 \\
\end{tabular}
\end{center}
\caption{ Experimental and computed melting transition temperatures.}
\label{table:tm}
\end{table}

\begin{figure}
\centering
\includegraphics[width=0.5\textwidth]{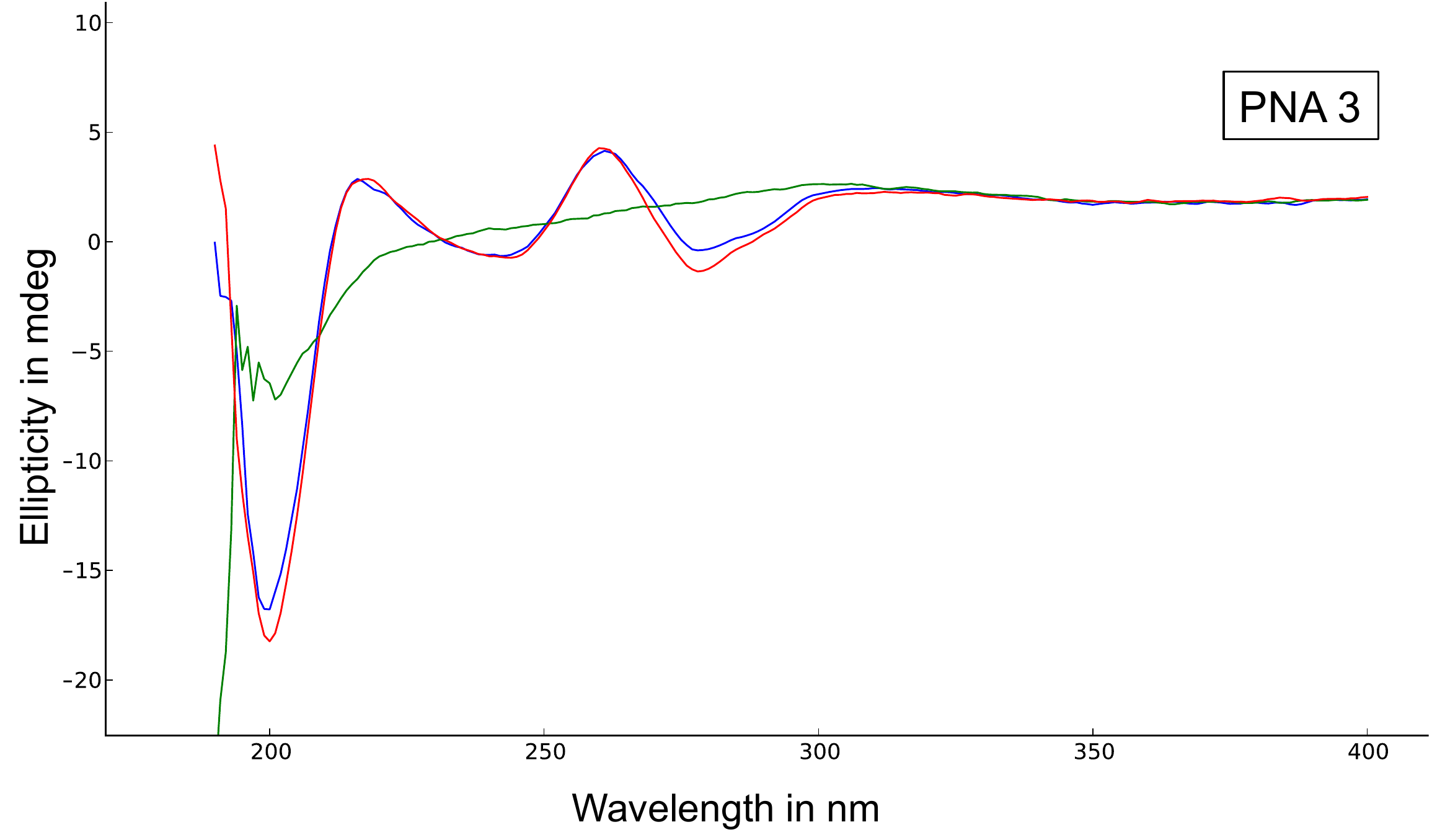}\hfill
\includegraphics[width=0.45\textwidth]{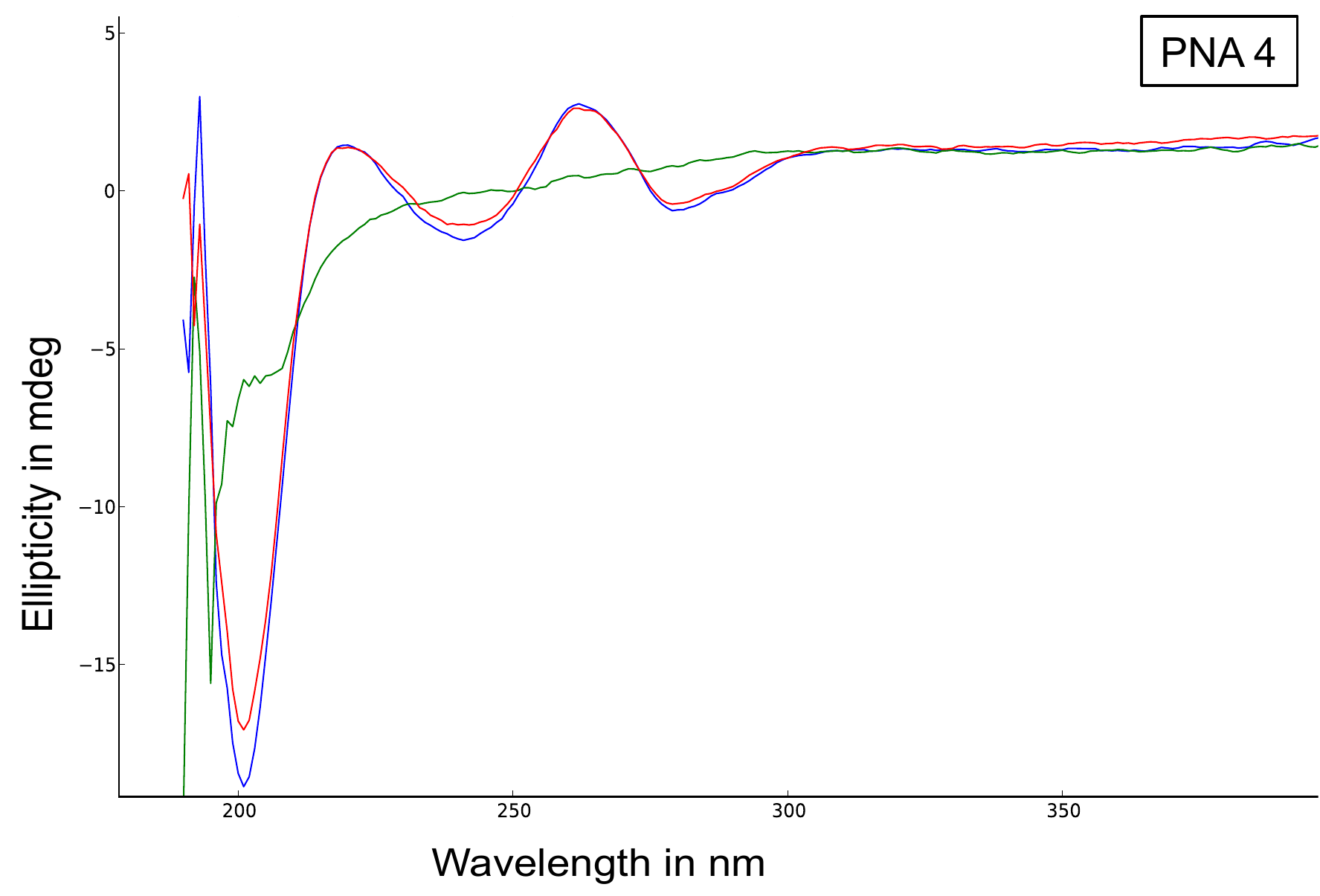}
\caption{Ellipticity curves of PNA \underline{3} (left) and PNA \underline{4} (right) at three temperatures.
Red and blue curves are ellipticity signal at 308\,K before and after denaturation respectively, green curve is ellipticity measured at 368\,K.
}
\label{cd_3_4}
\end{figure}

Full circular dichroism spectra of PNA \underline{3} and \underline{4} recorded at 308\,K and 368\,K are presented in figure \ref{cd_3_4}.
The ellipticity signals are very different at low and high temperatures.
At 308\,K (red curve) the complex signal observed for both PNAs is characteristic of a secondary structure existing in solution.
The dichroic signal varies according to the chirality of the molecule and of the secondary structure of the molecule.
This shows a pronounced Cotton effects, characteristic of a helix as show by Dragulescu-Andrasi et al\cite{DragulescuAndrasi:1970hq}.
This secondary structure is disrupted at high temperature (green curve), as at 368\,K the ellipticity signal presents much less structure and remains mostly flat, probably dominated only by the chirality of the backbone C$_\gamma$ carbons.
This absence of secondary structure at high temperature is characteristic of the melting of the duplex, as was already indicated by the hyperchromicity observed at high temperatures (see Supplementary Material figure S17).
The curves recorded at 308\,K before and after (blue curve) the high temperature denaturation are fully superimposed, indicating that the secondary structure of PNA in solution is fully reversible.
All these results are indicative of the formation of a chiral supramolecular organization, probably in the form of a helical secondary structure of the double strand.

Comparison of the overall shape of the CD spectra with the literature\cite{Tedeschi:2005kt}, and considering an antiparallel duplex, the maxima around 262 and 217\,nm, and the minima around 277, 238, and 200\,nm indicate that the formed duplex is probably a left handed helix.

\begin{figure}
\centering
\includegraphics[width=0.45\textwidth]{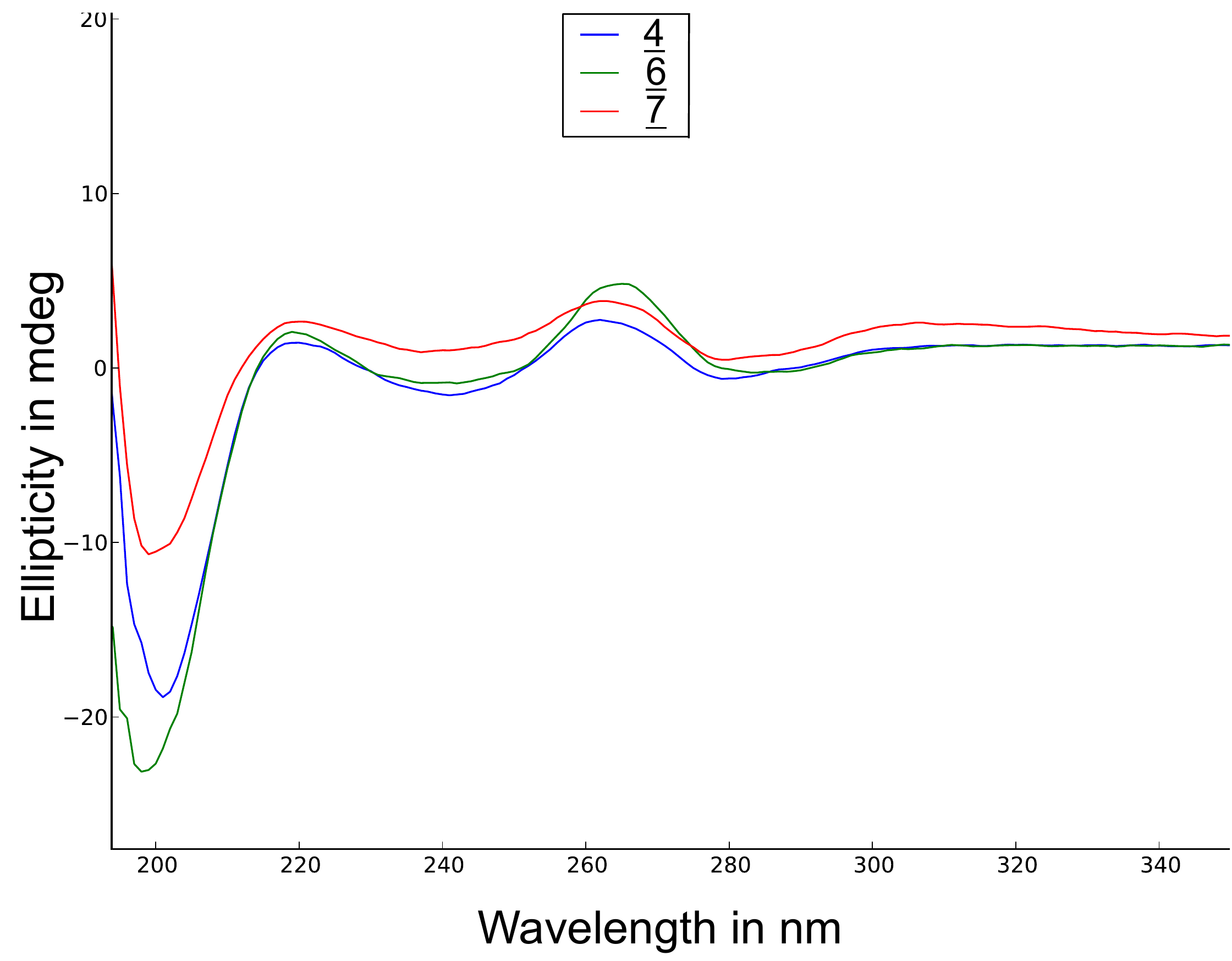}
\caption{Ellipticity curves of PNA \underline{4} (blue), PNA \underline{6} (green) and PNA \underline{7} (red) as a function of the wavelength.
PNA \underline{7} is duplex of PNA \underline{4} and PNA \underline{6}.}
\label{cd_4_6_7}
\end{figure}

In the same manner, ellipticity curves for PNA \underline{4}, \underline{6} and \underline{7} are presented in figure \ref{cd_4_6_7}.
All three structures display similar curves, indicating similar supramolecular organization.
However upon detailed analysis, we can observe that the maxima of the CD spectra occur at slightly different wavelengths, characteristic of slightly different structural organizations.

\section*{Conclusion} 
\label{sec:conclusion}

In conclusion, we have shown that PNA can adopt homo- and hetero-duplex conformations in solution through the formation of standard Watson-Crick as well as non-standard base pairing.
These duplexes are extremely stable and present melting transitions at temperatures higher than their nucleic acid homologues.
An extreme melting transition well above 100\,$^\circ$C was observed for a 10 bases complementary duplex, more than 40\,$^\circ$C higher than predicted for an RNA equivalent duplex.

PNAs duplexes with a chiral backbone present a marked chiral secondary structure, indicating that they are organized in supramolecular helices.
Circular Dichroism spectra indicate a common folding pattern for all studied structures, with nevertheless small differences depending on the details of the nucleobase sequence.


\section*{Acknowledgements} 
\label{sec:acknowledgements}
This work was financially supported by the NMRTEC company (Illkirch - France), and the Association Nationale de la Recherche et de la Technologie, CIFRE number 376/2010.
IGBMC is acknowledged for access to the NMR and CD spectrometers.
JV want to thank Yves Nomin\'e for his help with CD measurements.


\section*{Supplementary Data} 
\label{sec:supplementaray_data}

Supplementary data associated with this article can be found in the online version.
\begin{itemize}
	\item Document 1 : Figures S11--S14; NMR spectra of the different PNAs
    \item Document 2 : Programs and Figures S15--S35; Analysis of the Absorbance and CD data.
\end{itemize}


\section*{References} 

%







\section*{Supplementary material}

# Supplementary Materials - I
## Secondary structure and self pairing of gamma-PNA observed by NMR and CD

J.M.P. Viéville[a], S. Barluenga[b], N. Winssinger[b], M-A. Delsuc[c,]

[a]*Strasbourg University, Plateforme d'Analyse Chimique de Strasbourg Illkirch, 74 route du Rhin 67401 ILLKIRCH, FRANCE*
[b]*Department of Organic Chemistry, University of Geneva, Geneva CH1211, SWITZERLAND*
[c]*IGBMC, CNRS UMR 7104, 1 rue Laurent Fries BP10142, 67404 ILLKIRCH FRANCE*

**Abstract**

This documents presents additional spectra obtained on the different PNA molecules.

*Email address:* `delsuc@igbmc.fr` (M-A. Delsuc)

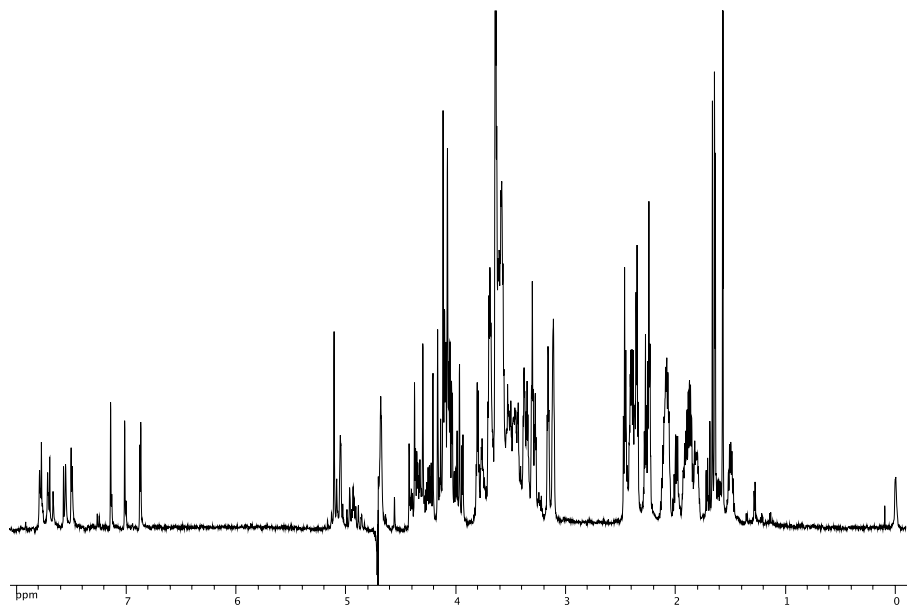

Figure S1: 1D $^1H$ NMR spectrum of PNA 1 at 700MHz with solvent presaturation. [1]=$1.56mM$, [Phosphate]=$25mM$, T=298K.

NMR spectrum of PNA 1 shows more aromatics protons than expected, it seems to have several methyls groups. This sample is not completely clean.

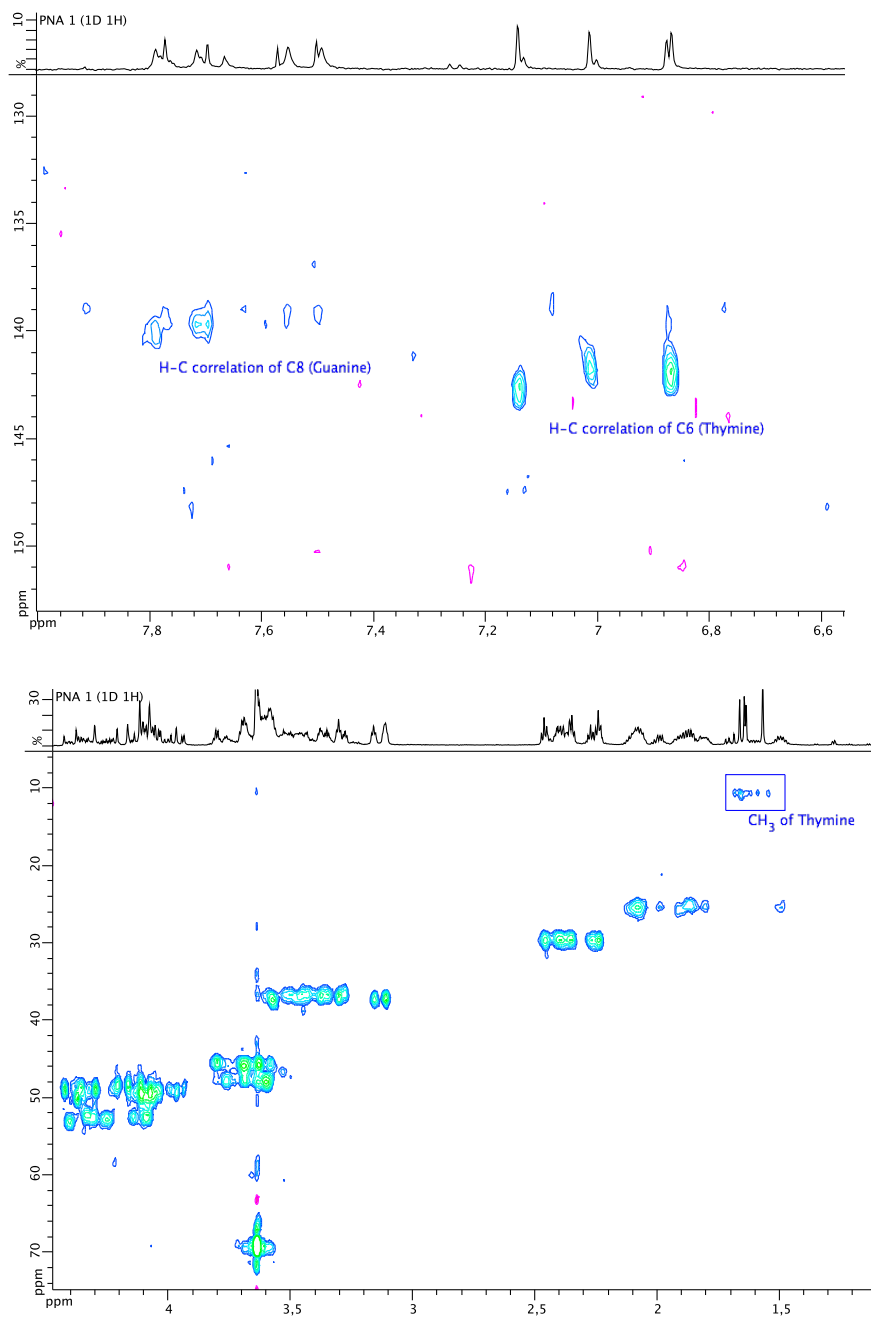

Figure S2: 2D $^{13}$C-HSQC NMR spectrum of PNA $\underline{1}$ at 700MHz. $[\underline{1}]=1.56mM$, [Phosphate]$=25mM$, T=298K.

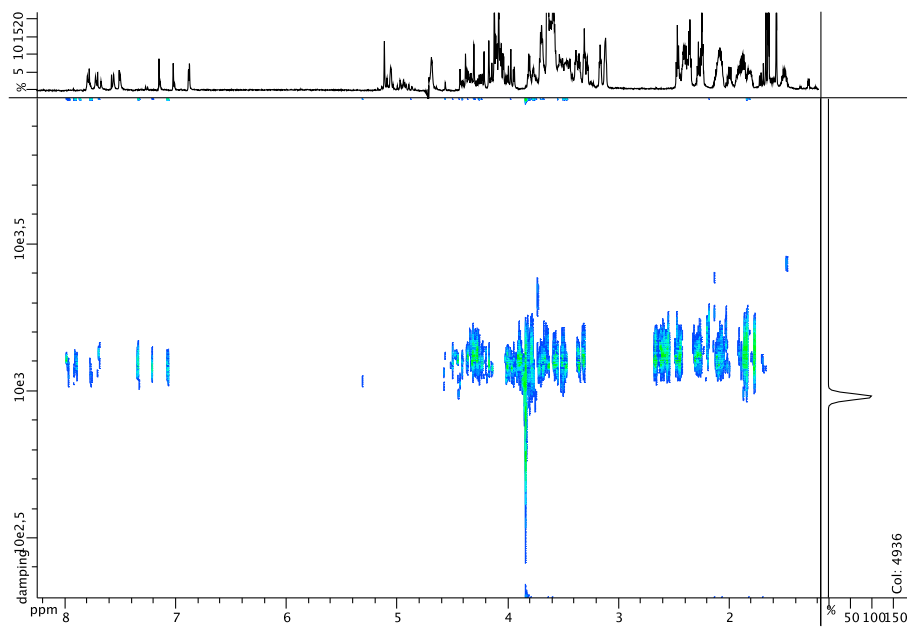

Figure S3: 2D DOSY NMR spectrum of PNA 1 at 700MHz with excitation sculpting. [1]=$1.56 mM$, [Phosphate]=$25 mM$, T=298K.

On the HSQC spectrum of PNA 1 (S2), we highlight the presence of two or three methyl groups. The PNA 1 is GGT made of, only one methyl of thymine should be present. Guanine groups are present on the HSQC experiment. As seen on the aliphatique area (3-5ppm), a mix of PNA could be present. The DOSY experiment shown in S3 doesn't confirm the presence of other PNA. It is not exclude to have other 3 nucleobases PNA with a similar molecular weight and apparent diffusion coefficient.

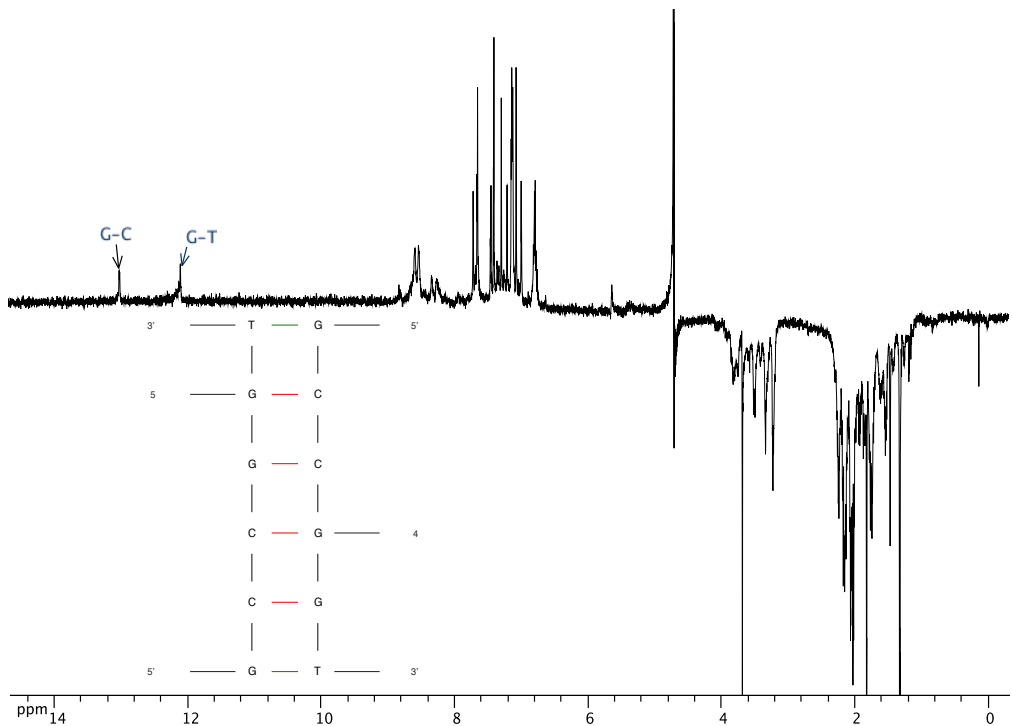

Figure S4: 1D $^1H$ `Jump & Return` NMR spectrum of PNA 2 at 700MHz. [2]=$880\mu M$, [Phosphate]=$25mM$, T=298K. The DINAMelt prediction for homodimerization is drawn.

PNA 2 should form homodimer as drawn in the insert of the picture S4. G-T and G-C base pairing are apparent on the `Jump & Return` NMR spectrum, indicating a self base pairing. Intensities of peak are non proportional to the proton number in the `Jump & Return` experiment as soon as they are far from the center of the window. Thus, the G-C imino protons peak appears weaker than the G-T imino proton peak which have twice a less number of protons.

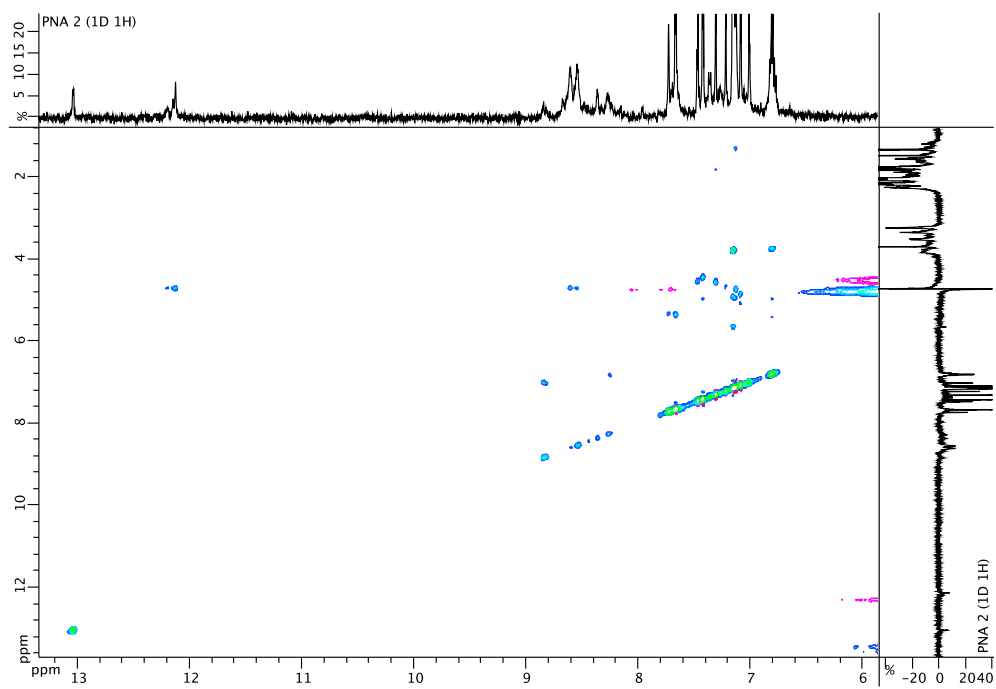

Figure S5: 2D $^1H-^1H$ `Jump & Return` NOESY NMR spectrum of PNA $\underline{2}$ at 700MHz. [2]=$880\mu M$, [Phosphate]=$25mM$, T=313K, mixing time= $250 msec$.

The NOESYjr NMR spectrum of PNA $\underline{2}$ does not highlight presence of cross peak between G-C imino proton and water.

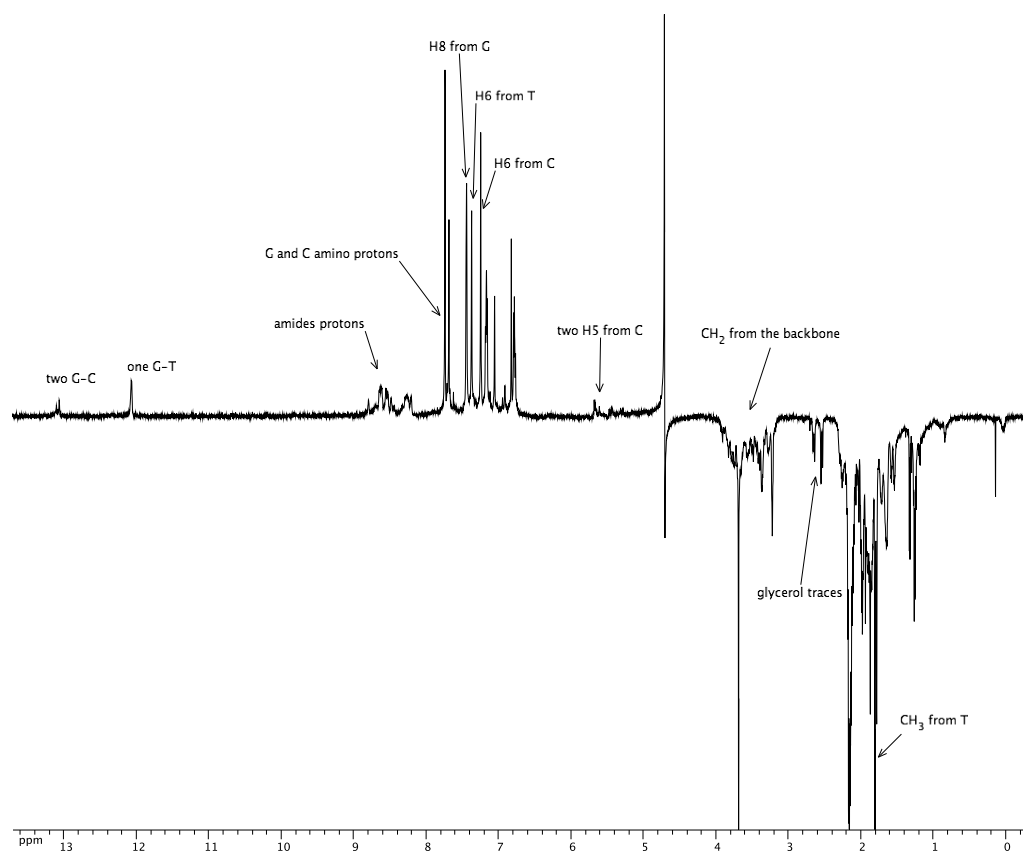

Figure S6: Partial assignment of PNA $\underline{3}$ at 700MHz. $[\underline{3}]=1.1mM$, [Phosphate]$=25mM$, T=298K.

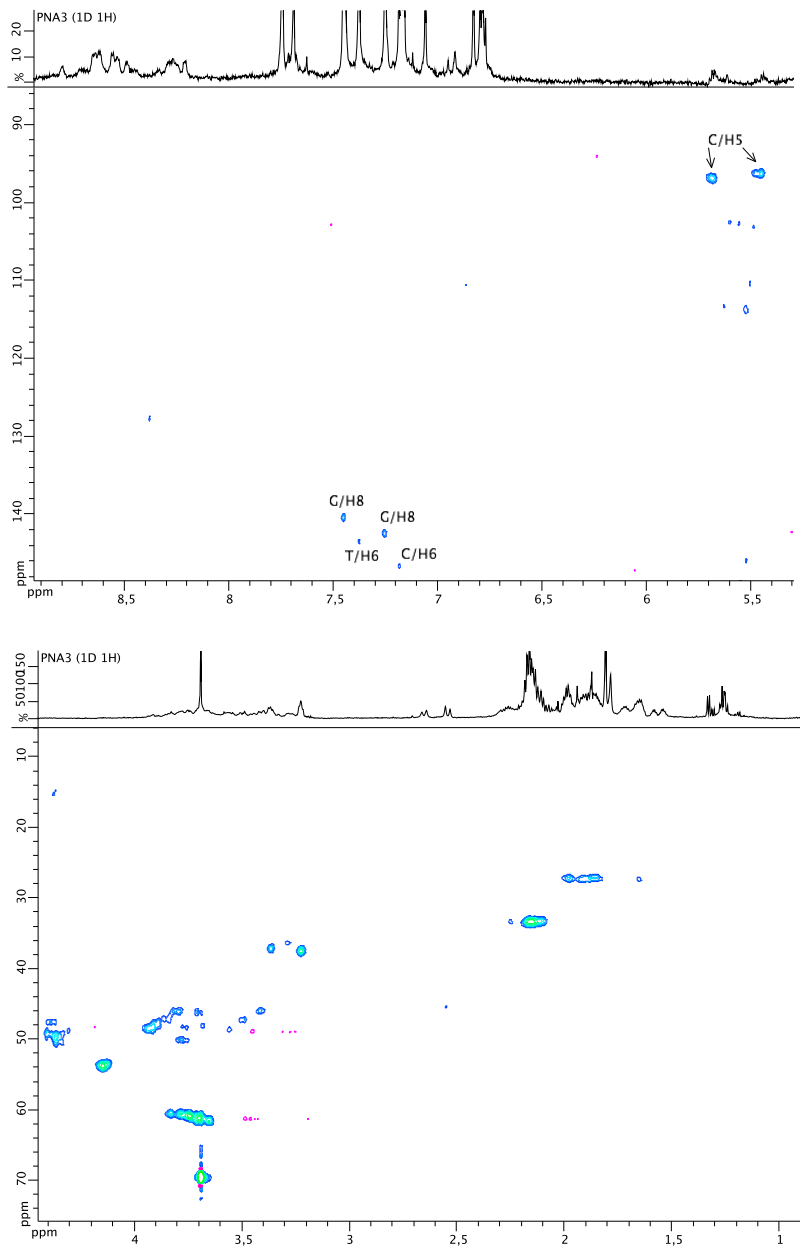

Figure S7: 2D $^{13}$C-HSQC NMR spectrum of PNA $\underline{3}$ at 700MHz. [$\underline{3}$]=$1.1 mM$, [Phosphate]=$25 mM$, T=303K. Acquisition time= 1 day.

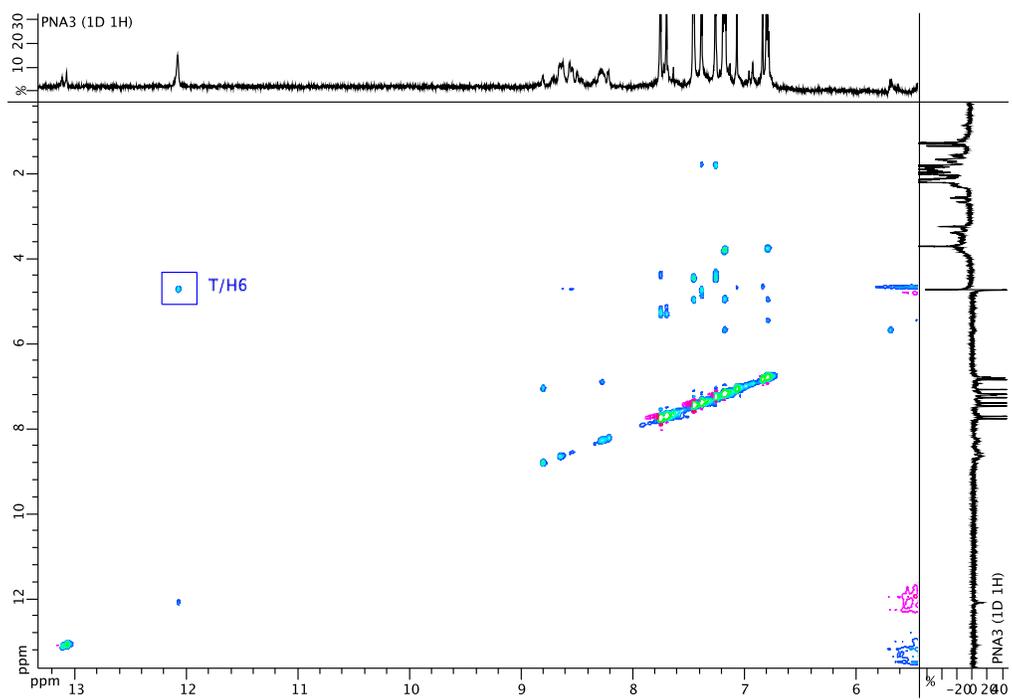

Figure S8: Zoom of 2D $^1H -^1H$ `Jump & Return` NOESY NMR spectrum of PNA $\underline{3}$ at 700MHz. [$\underline{3}$]=$1.1mM$, [Phosphate]=$25mM$, T=303K, mixing time= $250msec$, acquisition time= 1 day and 2 hours.

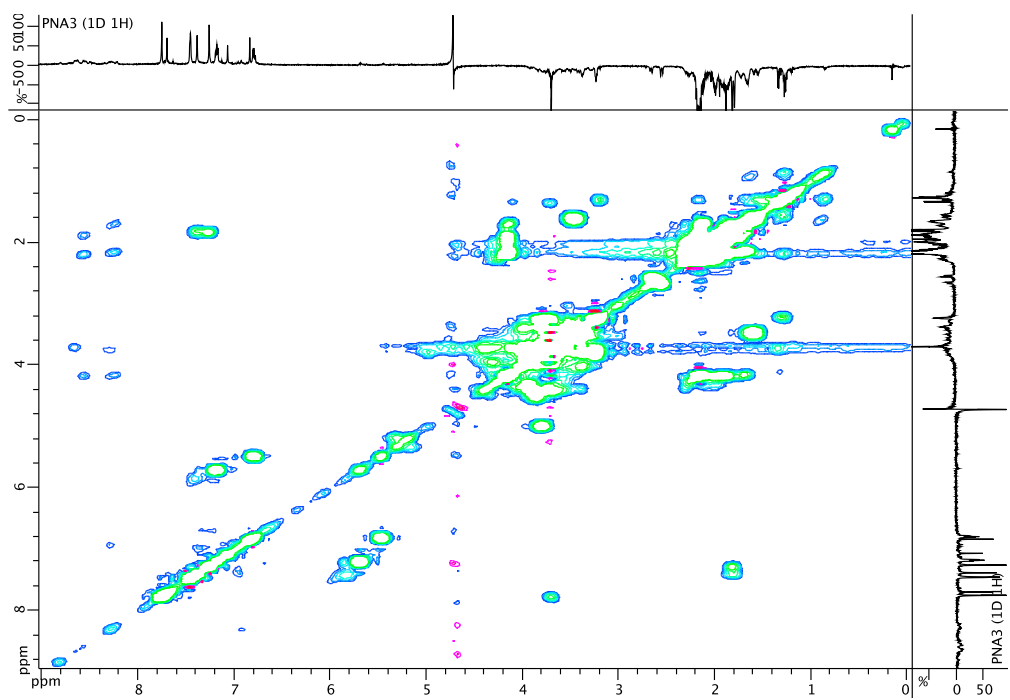

Figure S9: Zoom of 2D $^1H-^1H$ TOCSY NMR spectrum of PNA $\underline{3}$ at 700MHz. [$\underline{3}$]=1.1$mM$, [Phosphate]=25$mM$, T=303K, mixing time= 90$msec$, acquisition time= 19 hours.

23

Temperature variation of PNA $\underline{4}$ shows at least two G-C and two or three G-T imino protons, The G-T base pairing appears strong as G-T imimo protons signal remain visible at 318K.

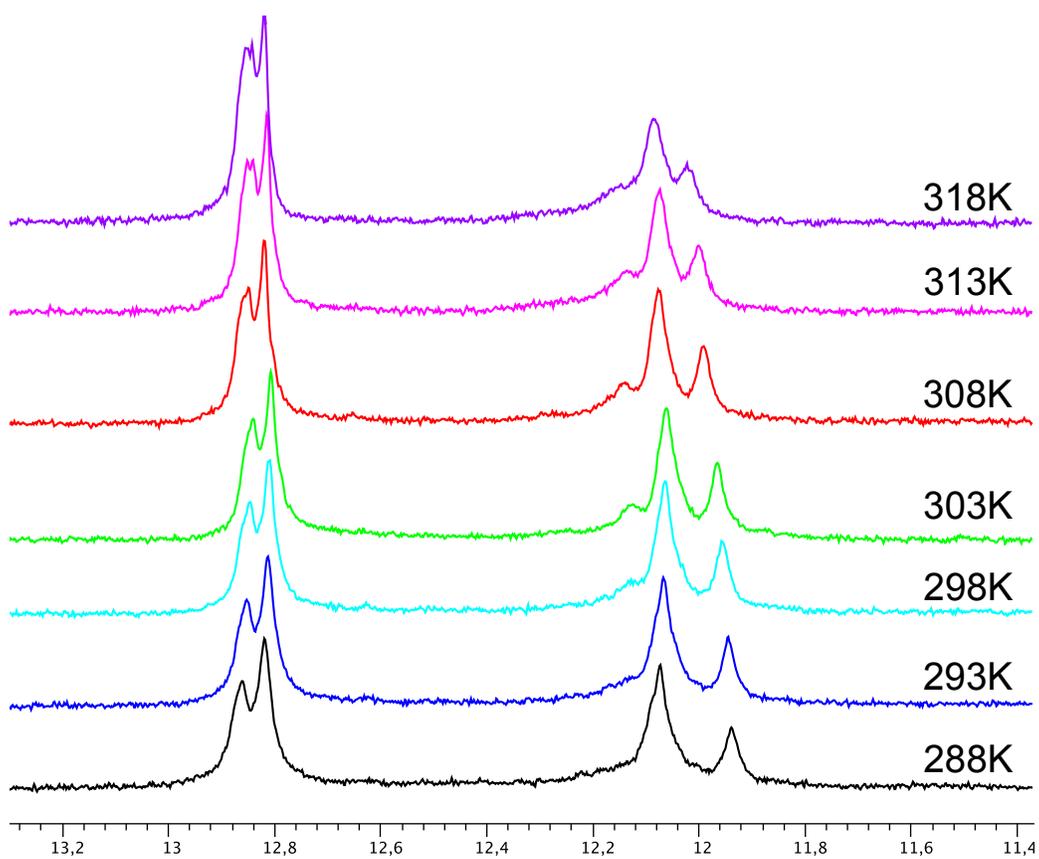

Figure S10: 1D $^1H$ `Jump & Return` NMR spectrum of PNA $\underline{4}$ at 700MHz. $[\underline{4}]=1.16mM$, [Phosphate]$=25mM$, variable temperature.

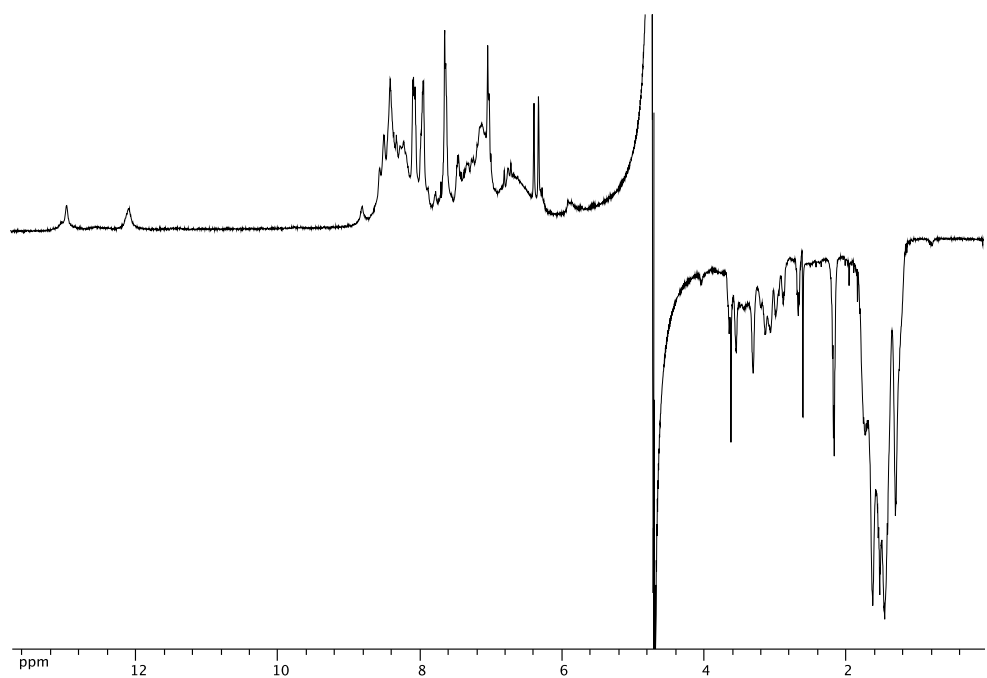

Figure S11: 1D $^1H$ `Jump & Return` NMR spectrum of PNA $\underline{5}$ at 500MHz. [$\underline{5}$]=$1.71mM$, [Phosphate]=$25mM$, T=298K.

In comparison the to PNA 4, the Biotin ended PNA5 shows one G-C base pairing and one large signal for a G-T base pairing. The G-T base pairing appears less stable as the G-T imino proton signal decrease at high temperature. This imino proton signal was still evident on PNA 4 at 318K, whereas this signal vanish at the same temperature in PNA 5, meaning the biotin impact on the folding of these PNA.

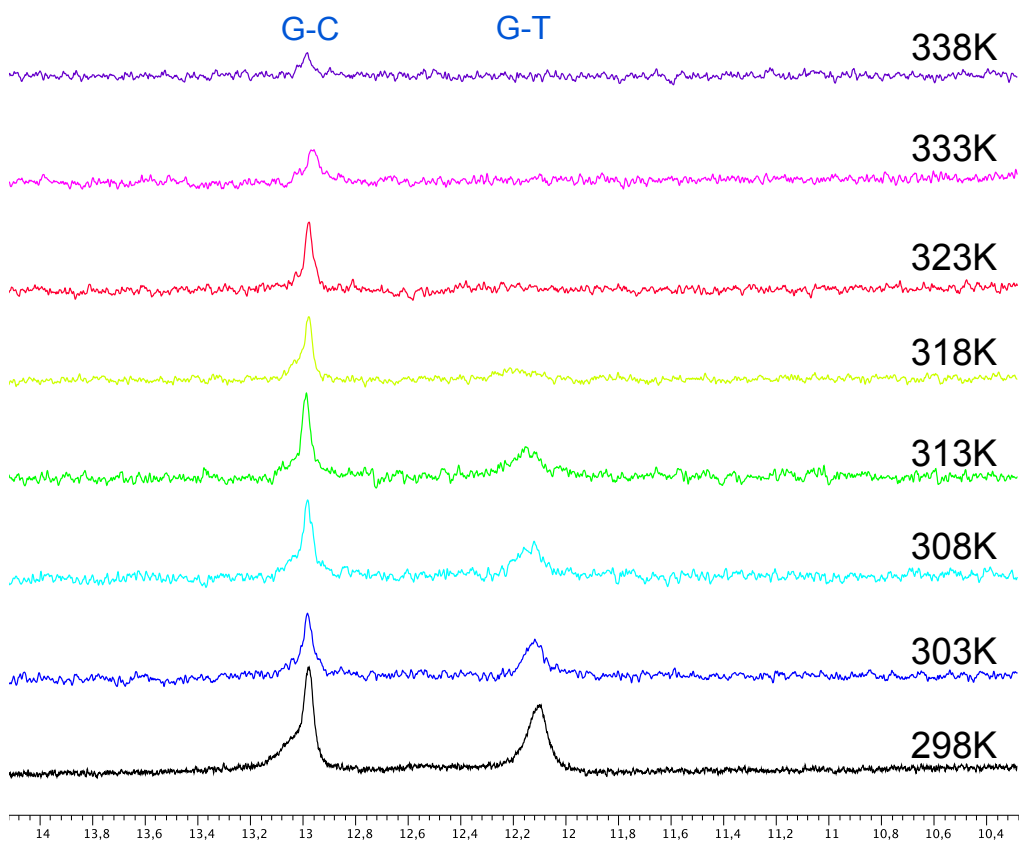

Figure S12: 1D $^1H$ `Jump & Return` NMR spectrum of PNA $\underline{5}$ at 500MHz. [$\underline{5}$]=$1.71mM$, [Phosphate]=$25mM$, variable temperature.

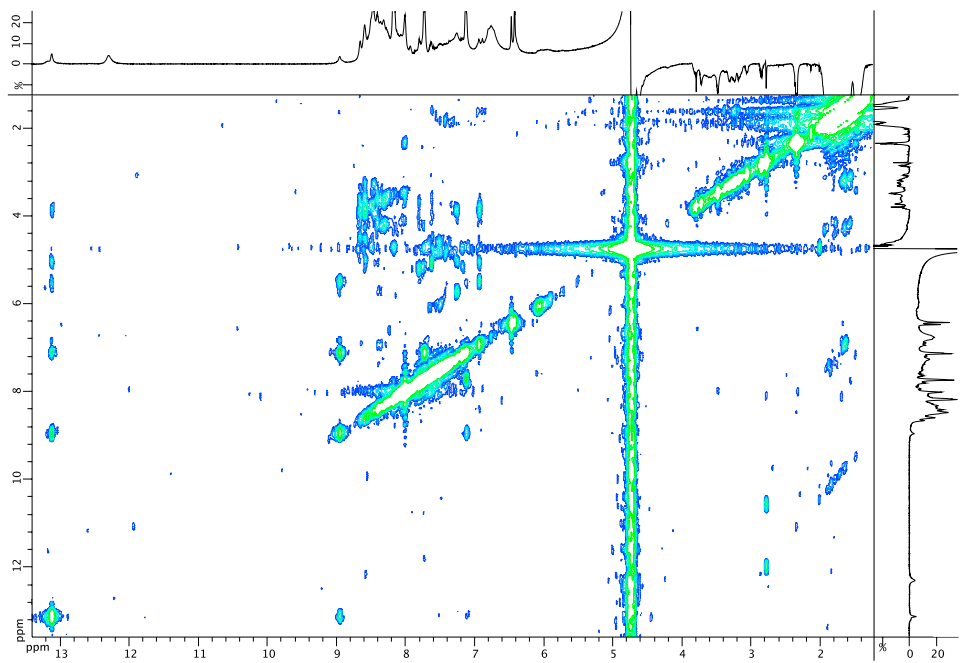

Figure S13: 2D $^1H\,-^1H$ `Jump & Return` NOESY NMR spectrum of PNA $\underline{5}$ at 700MHz. [$\underline{5}$]=$1.71mM$, [Phosphate]=$25mM$, T=313K, mixing time= $400msec$.

For the first time, the NOESY spectrum of PNA $\underline{5}$ shows correlation between one imino proton and protons from the backbone chain. Sadly, the complete assignement of this PNA remain impossible on this sample.

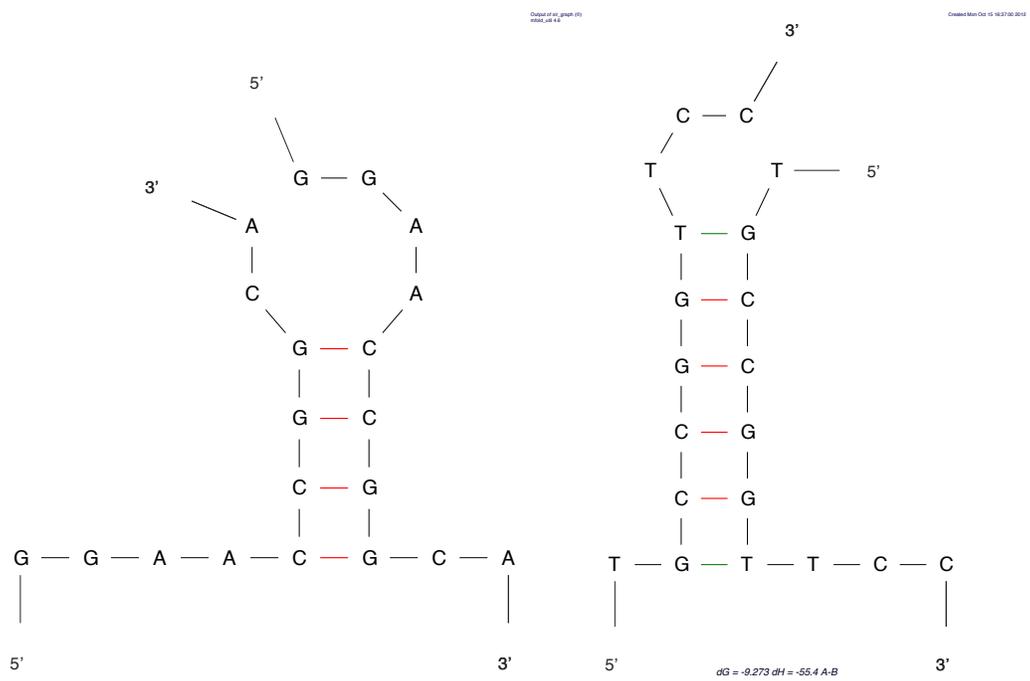

Figure S14: DINAMelt prediction of homodimerization of PNA 5 and 4 (right) and 6 (left).

# Melting_analysis

August 12, 2015

# 1 Duplex formation and secondary structure of $\gamma$-PNA observed by NMR and CD *Supplementary Material*

# 2 Content

This file present the program used to determine the melting temperatures $T_m$ of PNA molecules, from absorbance and CD data extracted acquired on Jasco spectrometer.

The document is in two parts

- the fitting program itself,
- the results obtained when applying the program to experimental data provided in another file-set.

This Program are provided under the Licence CeCILL 2.1

This program HAS NOT been tested intensively, it is believed to do what it is supposed to do, Howver, you are welcome to check it on your own data.

- Author : M-A Delsuc (madelsuc@unistra.fr)
- Date : August 2015
- Version : 2.0 (Multiexp and error bars was added from version 1.0, also some modification have been made to make the prgm easier to use )

The program and the datasets are freely available at https://github.com/delsuc/Melting-curve-analysis

# 3 The program

This program is a general fitting program, meant to analyze melting curves, either from Absorption or CD data. It has been used here to analyze the PNA data presented in the associated work, but can certainly be used on any other data.

It is in three parts :

- reading the Jasco files
- This part in independent from the other part, and can be extended to adapt your own set-up, the main entry point is `load()` .
- JASCO csv and txt files are programmed
- the `FrenchCoding` flag allows to access the files on a french version of Windows,
- defining a theoretical melting curve, used for fitting
- this is supposed to be general to any copperative melting mechanism
- The program itself that read files, fit the melting curve and display the results.

The program comes in two versions:

- `Monoexp()` permits to fit one melting experiment, measured at one given wavelength
- `Multiexp()` permits to fit several melting experiments, measured at several wavelengths

```python
In [1]: #first, set-up the python scene
        import sys
        import codecs
        import numpy as np
        import matplotlib as mpl
        import matplotlib.pyplot as plt
        from scipy.optimize import curve_fit
        Debug = False

        # these seaborn is optional, used for a nice setup
        try:
            import seaborn as sns
            current_palette = sns.color_palette()
        except ImportError:
            pass

        %matplotlib inline
        mpl.rc("figure", figsize=(12, 6))

        from IPython.display import HTML, display_html
        TheFig = 15
        def Fig(caption=None):
            global TheFig
            st = "<b>Figure S%d</B>"%TheFig
            TheFig += 1
            if caption:
                st += " : <i>%s</i>"%caption
            display_html(HTML(st))
```

## 3.1 utilities to read the Jasco files

**Remark** Our JASCO system is running on a french version of Windows. Files are coded with the `latin_1` coding, depending on your set-up, you may need to change this parameter if you experiment coding errors while reading the file.

On the french system, decimal numbers are coded with a comma for decimal : `33,2` instead of `33.2`

Finally the `.csv` files on the french system are coded with a semi colon to separate the field, so a line in a CSV file appears as :

`1; 33,2; 64,5`

instead of

`1, 33.2, 64.5`

This is taking care of with the `FrenchCoding` variable set to `True`

```python
In [2]: # adapt FileCoding to the ASCII coding used in the JASCO text files
        # exemples are
        # "cp1252" (Western Europe)   "latin_1" (France and other)   "shift_jis" (japanese)   "utf_8"
        # check https://docs.python.org/2.7/library/codecs.html#standard-encodings
        #FileCoding = "utf_8"
        FileCoding = "latin_1"

        # set to FrenchCoding to False if you the cvs and txt files are not using the french coding
        # (, for decimal point and ; for field separator )
```

```python
#FrenchCoding = False
FrenchCoding = True

#read files
def loadcsv(fich):
    """
    reads a csv jasco file
    returns data, meta
    where  data as a numpy array
           meta as a dictionary key:value
    """
    data = []
    meta = {}
    xydata = False
    for lin in codecs.open(fich, "rb", FileCoding):  # the codec may be different on your set-up
        if FrenchCoding:
            lin = lin.replace(',','.')
            linspl = lin.strip().split(';')
        else:
            linspl = lin.strip().split(',')
        if linspl == ['']:         # marks end of data
            xydata = False
        if xydata:                 # reading data
            flin = [ float(i) for i in linspl]
            data.append(flin)
        elif len(linspl) >1 :      # ou pas    un champs a recuperer
            meta[linspl[0]] = ";".join(linspl[1:])
        if linspl == ["XYDATA"]:   # marks beginning of data
            xydata = True
    return np.array(data), meta

def loadtxt(fich):
    """
    reads a text jasco file
    returns data, meta
    where  data as a numpy array
           meta as a dictionary key:value
    """
    data = []
    meta = {}
    xydata = False
    for lin in codecs.open(fich, "rb", FileCoding): # codec may vary
        if FrenchCoding:
            lin = lin.replace(',','.')   # This is for french coding on windows !
        lin = lin.strip()
        linspl = lin.split('\t')
        if linspl == ['']:         # marks end of data
            xydata = False
        if xydata:                 # read data
            flin = [ float(i) for i in linspl]
            data.append(flin)
        elif len(linspl) >1 :      #  if meta
            meta[linspl[0]] = "\t".join(linspl[1:])
        if linspl == ["XYDATA"]:   # marks beginning of data
```

```python
            xydata = True
        return np.array(data), meta

    def load(fich):
        "read jasco file, txt or csv based on extension .csv"
        try:
            if fich.endswith(".csv"):
                return loadcsv(fich)
            else:
                return loadtxt(fich)
        except:   # generic error message.
            if FrenchCoding:
                decimal = ","
            else:
                decimal = "."
            print """
            The data file could not be read
            please check
            - file name : %s
            - file coding : %s
            - decimal coding : %s
            """%(fich, FileCoding, decimal)
            sys.exit()

    def decoupe(d):
        "extract columns from data array"
        ldo = d[:,0]
        cd = d[:,1]
        volt = d[:,2]
        absorb = d[:,3]
        return (ldo, cd, volt, absorb)

    def readvalues(fich, tampon=None, lincor=False):
        """
        read file and return cd and abs
        if tampon is a file, will corrected for buffer
        if lincor is True, a linear component is removed
        returns lambda_array, absorb_array, cd_array, remark_text
        """
        d,meta = load(fich)
    #    print meta['Comment']
        (ldo, cd, volt, absorb) = decoupe(d)
        if tampon:
            dtamp,metatamp = load(tampon)
            (a, b ,c , absorb_tamp) = decoupe(dtamp)
            absorb = absorb-absorb_tamp
            rem = " - corrected"
        else:
            rem = ""
        if lincor:       #eventually correct for linear component evaluated on first points
            zone = int(0.1*len(ldo))
            fit = np.polyfit(ldo[0:zone], absorb[0:zone], 1)
            absorb = absorb-np.polyval(fit, ldo)
        return ldo, absorb, cd, rem
```

```python
    def plotcd(fich, tampon=None, lincor=False, label=None):
        "draws cd from file:fich with label,  with eventual corrections"
        ldo, absorb, cd, rem = readvalues(fich, tampon=tampon, lincor=lincor)
        if label is None:
            label = fich+rem
        plt.plot(ldo, cd, label=label)
        plt.legend()
        plt.title(fich)

    def plotabs(fich, tampon=None, lincor=False, label=None):
        "draws absorbance from file:fich with label,  with eventual corrections"
        ldo, absorb, cd, rem = readvalues(fich, tampon=tampon, lincor=lincor)
        if label is None:
            label = fich+rem
        plt.plot(ldo, absorb, label=label)
        plt.legend()
        plt.title(fich)
```

## 3.2 melting curve definition,

used for fitting, implements the hypoerbolic-tangent model, see for instance : M. YA. AZBEL, "DNA sequencing and melting curve" *P.N.A.S.* **76** (1) pp.101-105 (1979)

```python
In [3]: #T_m fitter
        def Tmfunc(T, Io, Ie, Tm, beta):
            """
            draw a theoretical fusion curve
            T: temperature range
            Io Ie : value at low and high temp
            Tm: fusion temperature
            beta: cooperativity parameter, proportional to the inverse of the temperature range over wh
            """
            if Debug:
                print "T",T,"\n args",(Io, Ie, Tm, beta)
            tred =   beta*(T-Tm)
            return Io + 0.5*(Ie-Io)*(1+np.tanh(tred))
        #and plot theoretical curve
        x = np.linspace(0,100,100)
        plt.plot(x, Tmfunc(x, 5,10,60,0.1), label=r"$\beta=0.1$")
        plt.plot(x, Tmfunc(x, 5,10,60,0.07),label=r"$\beta=0.7$" )
        plt.plot(x, Tmfunc(x, 5,10,60,0.05), label=r"$\beta=0.05$" )
        plt.xlabel("Temperature ($^\circ C$)")
        plt.ylabel("A.U.")
        plt.legend(loc=0)
        Fig(r"Theoretical melting curves for a $T_m$ of $60^\circ C$ with varying $\beta$")
```

Figure S15 : Theoretical melting curves for a $T_m$ of $60^\circ C$ with varying $\beta$

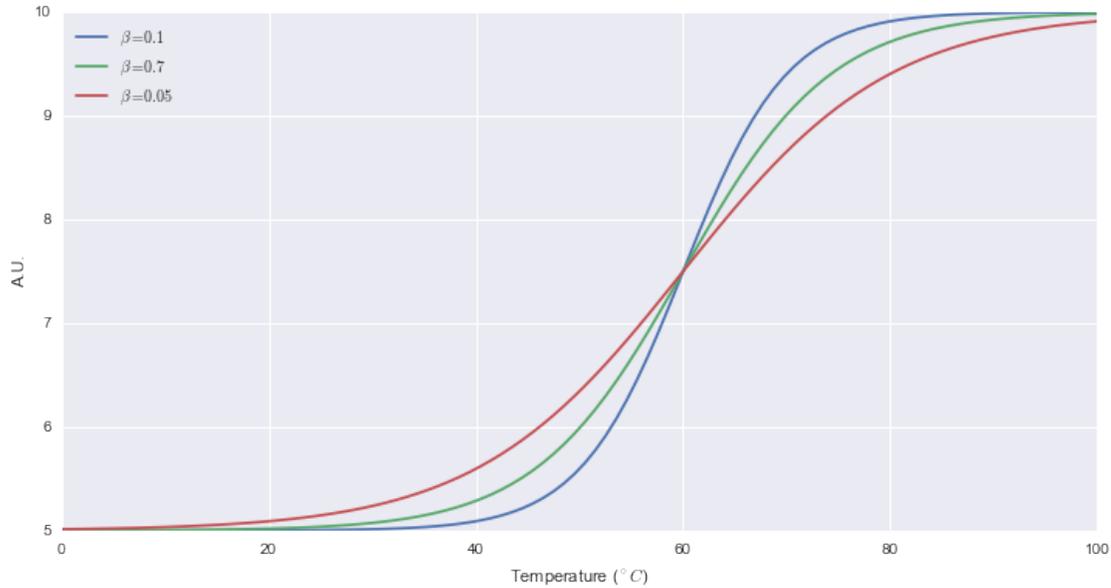

## 3.3 the fitting program

The class `Monoexp()` defines an object which loads the various data of a temperature run followed by CD/Absorbance at one wavelength.

To use, initialize the various attributes (parameters) (check `__init__` for documentation), attributes are :

- BASE : the base directory which holds the files
- fnom : the generic name for the file
- freq : the frequency (wavelength) on wich to apply the measure
- cell : the cell to follow
- tampon : None, or the cell of the buffer
- meth : meth is either "ABS" or "CD"
- nom : used for label in plots
- extension : ".txt" or ".csv"

then use either :

- `plot()` : plots the curve
- `analyze()` : plot and fit melting curve, determines $T_m$ and $\beta$

In [4]: ```
##### Create the programs that compute and display results
class Monoexp(object):
    """
    This class defines an object which loads the various data which describe one experiment
    of a temperature run followed at one wavelength by CD/Absorbance.

    To use,
     - initialize the various parameters (check '__init__()' function for detailled documentati
    then use either :
     - plot() : plots the curve
     - analyze() : plot and fit melting curve
    """
```

```python
    def __init__(self):
        """
        creates the object
        The file name will be built from these parameters with self.full()
        """
        self.BASE = "./"            # the base directory which holds the files
        self.fnom = "filename-"     # the generic name for the file
        self.freq = 217             # the frequency to measure
        self.cell = 2               # the cell to follow
        self.tampon = None          # None, or the cell of the buffer
        self.meth = "abs"           # meth is either "ABS" or "CD"
        self.nom = None             # used for label in plots
        self.extension = ".txt"     # ".txt" or ".csv"

    def ls(self):
        "utility to print all possible files, using the name preset"
        import glob
        fn = "%s/%s*%s"%(self.BASE, self.fnom, self.extension)
        print "List of files :"
        print "\n".join( glob.glob(fn) )
    @property
    def full(self):
        "the filename"
        return "%s/%s%dnm-Cell %d%s"%(self.BASE, self.fnom, self.freq, self.cell, self.extension)
    @property
    def tfull(self):
        "the filename of the buffer"
        if self.tampon is not None:
            return "%s/%s%dnm-Cell %d%s"%(self.BASE, self.fnom, self.freq, self.tampon, self.ext
        else:
            return None
    @property
    def title(self):
        "title for plot"
        if self.meth.lower() == 'cd':
            return 'CD over Temperature'
        elif self.meth.lower() == 'abs':
            return 'Absorbance  over Temperature'
    @property
    def label(self):
        "label for plot"
        if self.nom is None:
            self.nom = self.cell
        return 'PNA %s - $\lambda=%d\,nm$'%( self.nom, self.freq)
    def fit(self, Tm=60.0, beta=0.1, Io=None, Ie=None):
        """
        fit and plot result, initial estimate can be given to help the fit

        uses all current attributes (see __init__) to determine which dataset is to be fitted

        results are stored in the self?Results dictionary
        """
        global Debug
        ldo, absorb, cd, rem = readvalues(self.full, tampon=self.tfull)
```

```python
        self.ldo = ldo
        if self.meth.lower() == "abs":
            y = absorb
        elif self.meth.lower() == "cd":
            y = cd
        # set-up initial fit values
        if Io is None:
            Io = y[0]
        if Ie is None:
            Ie = y[-1]
        try:
            popt, pcov = curve_fit(Tmfunc, ldo, y, p0=[Io, Ie, Tm, beta] ) # fit
        except RuntimeError:
            print self.label, "curve could not be fitted"
            self.Results = None
        else:
            perr = 2*np.sqrt(np.diag(pcov))
            self.Results = {
                'Io':popt[0],
                'Ie':popt[1],
                'Tm':popt[2],
                'beta':popt[3],
                'Tm_errorbar':perr[2],
                'beta_error':perr[3] }# store monofit
            self.residu = y-Tmfunc(ldo, *popt)
            self.norm = np.sqrt(np.sum(self.residu**2))
            if Debug: print pcov
            print "%s  Tm: %.1fC +/- %.2f  beta: %.2f  chi2:%.3f"%(self.label, popt[2], perr[2]
    def showfit(self):
        "plot the result of the fit"
        try:
            R = self.Results
        except:
            raise Exception("fit is not performed yet")
        if R is not None: # None indicates fit not converged
            plt.plot(self.ldo, Tmfunc(self.ldo, R["Io"], R["Ie"], R["Tm"], R["beta"]), 'k--') #
            mn = min( R["Io"], R["Ie"] )
            mx = max( R["Io"], R["Ie"] )
            if R["Tm"] < 100.0 and R["Tm"] > 0.0:    # do not plot out-of-range results
                plt.plot([R["Tm"],R["Tm"]], [mn,mx], 'k:')

    def plot(self):
        "plot the experimental curve"
        if self.meth.lower() == "abs":
            self.abs()
            plt.ylabel("A.U.")
        elif self.meth.lower() == "cd":
            self.cd()
            plt.ylabel("Ellipticity - millideg")
        plt.legend(loc=0)
        plt.title(self.title)
        plt.xlabel("Temperature ($^\circ C$)")
    def analyze(self, Tm=60.0, beta=0.1, Io=None, Ie=None):
        "fit (using initial guess if provided)  and plot"
```

```python
            self.plot()
            self.fit(Io=Io, Ie=Ie, Tm=Tm, beta=beta)
            self.showfit()
        def abs(self):
            "read absorbance values"
            ldo, absorb, cd, rem = readvalues(self.full, tampon=self.tfull)
            plt.plot(ldo, absorb, label=self.label)
        def cd(self):
            "read CD values"
            ldo, absorb, cd, rem = readvalues(self.full, tampon=self.tfull)
            plt.plot(ldo, cd, label=self.label)
```

## 3.4 second method

The class `Multiexp()` is very close to `Monoeexp`, the only difference, is that it allows to fit several measures at different wavelength, to the same $T_m$ value

It is to initialize in the same manner than Monoexp, It has an additional attribute :

- multifreq : a list of the wavelengths to use in the fitting/plotting operations

then use either :

- `plot()` : plots the curve
- `analyze()` : plot and fit all the curves named by multifreq to a same $T_m$

```python
In [5]: def TmMfunc(T, *args):
            """
            Equivalent to Tmfunc, bur for several wavelengths, using the same Tm and beta
            everything is flattened into 1D arrays (parameters, and return values)

            T: temperature range
            N, Io, Ie, Tm, beta = *args
            Tm: fusion temperature
            beta: cooperativity parameter
            N: number of curve to analyze,
            then for each curve
                Io Ie : value at low and high temp
            """
            global Ncurves, Debug  # Ncurves contains the number of curves to fit
            if Debug:
                print "T",T,"\n args",args
            N = Ncurves
            Io = args[0:N]
            Ie = args[N:2*N]
            Tm = args[-2]
            beta = args[-1]
            r = []
            for i in range(len(Io)):
                tred = beta*(T[i]-Tm)
                r.append(Io[i] + 0.5*(Ie[i]-Io[i])*(1+np.tanh(tred)))
            return np.array(r).flatten()
        class Multiexp(Monoexp):
            def __init__(self):
                super(Multiexp, self).__init__()
                self.multifreq = []
```

```python
def fit(self, Tm=60.0, beta=0.1, Io=None, Ie=None):
    """
    fit and plot result, initial estimate can be given to help the fit
    values are given for initial estimate

    uses all current attributes (see __init__) to determine which dataset is to be fitted

    results are stored in the self.Results dictionary
    """
    global Ncurves   # Ncurves contains the number of curves to fit
    y = []
    self.ldo = []
    for self.freq in self.multifreq:
        ldo, absorb, cd, rem = readvalues(self.full, tampon=self.tfull)
        self.ldo.append(ldo)
        if self.meth.lower() == "abs":
            y.append(absorb)
        elif self.meth.lower() == "cd":
            y.append(cd)
    # set-up initial fit values
    if Io is None:
        Io = [i[0] for i in y]
    if Ie is None:
        Ie = [i[-1] for i in y]
    y = np.array(y).flatten()
    self.ldo = np.array(self.ldo)
    Ncurves = len(Io)
    p0 = Io + Ie + [Tm] + [beta] # all parameters are concatenated
    r = (TmMfunc(self.ldo, *p0 ) - y)
    if Debug:
        print self.ldo.shape, y.shape, len(p0)
        print TmMfunc(self.ldo, *p0 ).shape
        print p0
        print r
    try:
        popt, pcov = curve_fit(TmMfunc, self.ldo, y, p0=p0, maxfev = 6000)
    except RuntimeError:
        popt = p0
        perr = None
    else:
        print self.fitlabel, "curve could not be fitted"
        if Debug: print popt
        self.residu = y-TmMfunc(self.ldo, *popt)
        self.norm = np.sqrt(np.sum(self.residu**2))
        perr = 2*np.sqrt(np.diag(pcov))
        print "%s  Tm: %.1fC +/- %.2f   beta: %.2f   chi2:%.3f"%(self.fitlabel, popt[-2], pe
    if Debug:
            print popt
            self.residu = y-TmMfunc(self.ldo, *popt)
            self.norm = np.sqrt(np.sum(self.residu**2))
    if perr is not None:
        self.Results = {
            'N':Ncurves,
            'Io':popt[0:Ncurves],
```
10

```
                    'Ie':popt[Ncurves:2*Ncurves],
                    'Tm':popt[-2],
                    'beta':popt[-1],
                    'Tm_errorbar':perr[-2],
                    'beta_error':perr[-1] }# store monofit
            else:
                self.Results = None

    def showfit(self):
        "plot the result of the fit"
        try:
            R = self.Results
        except:
            raise Exception("fit is not performed yet")
        if self.Results is not None:
            for i in range(R['N']):
                plt.plot(self.ldo[i], Tmfunc(self.ldo[i], R["Io"][i], R["Ie"][i], R["Tm"], R["b
            mn = min( min(R["Io"]), min(R["Ie"]) )
            mx = max( max(R["Io"]), max(R["Ie"]) )
            if R["Tm"] < 100.0 and R["Tm"] > 0.0:     # do not plot out-of-range results
                plt.plot([R["Tm"],R["Tm"]], [mn,mx], 'k:')
    def plot(self):
        for self.freq in self.multifreq:
            super(Multiexp, self).plot()
    @property
    def fitlabel(self):
        "label for fit"
        if self.nom is None:
            self.nom = self.cell
        return 'PNA %s'%( self.nom)
```

# 4 Experimental Results

A directory called PNA_DATA/ should be available along with this program. It contains a set of files holding the data acquired on the various PNA studied.

The experiments were acquired on by Absorbance and CD on a Jasco J-815 spectropolarimeter using 1mm path length 110 quartz UV cells. Each run was performed first by heating, the temperature varying from 308 K to 368 K with a 1 K/mn variation rate, and then by a cooling experiment performed at the same rate.

Each file contains one measure on one PNA, at one wavelength, on a temperature series. The files are distributed among several folders corresponding to different experimental runs. The name of the file codes for the sample, the wavelength and the temperature variation (going up or down)

## 4.1 PNA 2 and 3

melting followed by absorbance and CD

- buffer is in 1
- PNA 2 in 2
- PNA3 in 3

### 4.1.1 CD Spectrum

First we plot the CD spectrum :

```
In [6]: plotcd(fich="PNA_DATA/Dichro_130627/PNA8xx_T35a_1mm_130627-3.txt",
               tampon="PNA_DATA/Dichro_130627/PNA8xx_T35a_1mm_130627-1.txt", lincor=False, label="PNA 3
        plotcd(fich="PNA_DATA/Dichro_130627/PNA8xx_AftDenatT35a_1mm_130627-3.txt",
               tampon="PNA_DATA/Dichro_130627/PNA8xx_T35a_1mm_130627-1.txt", lincor=False, label="PNA 3
        Fig(r"CD spectra of PNA 3 at 308K before and after the variation temperature")
```

Figure S16 : CD spectra of PNA 3 at 308K before and after the variation temperature

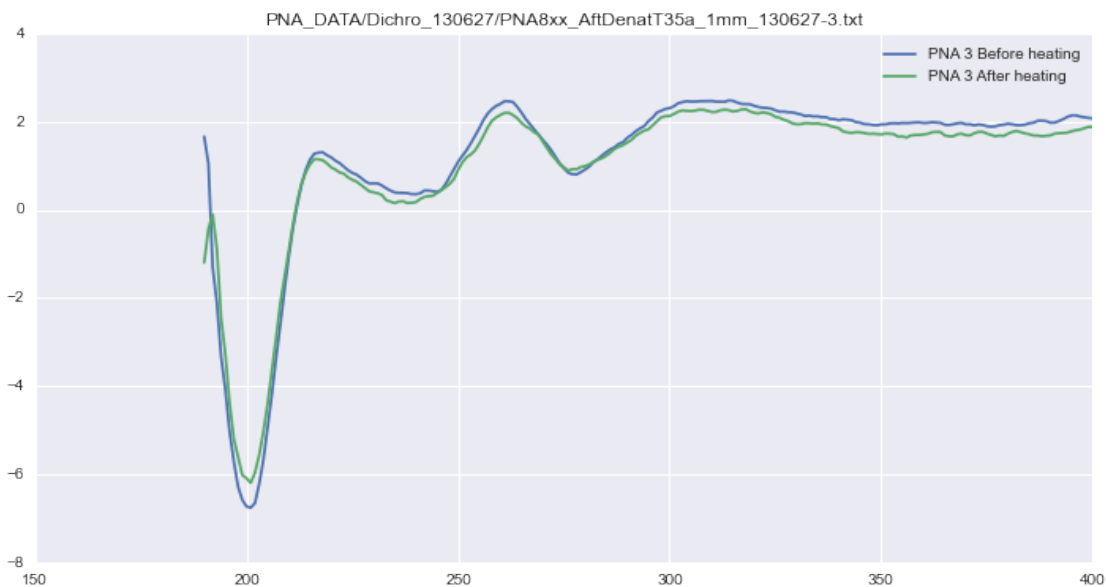

### 4.1.2 Temperature variation - going-up

1°C / min

The following wave-lengths were recorded : 200nm 217nm 230nm 260nm 310nm 380nm

Results are shown for **Absorbance** signal, and for **CD** signal for PNA-3 (PNA-2 is non chiral)

Not all frequencies give fittable results.

First we can use `Monoexp()` to fit the best curve.

```
In [7]: f = Monoexp()
        f.BASE = "PNA_DATA/Dichro_130627"
        f.fnom = "PNA8xx_Denat_130627-1-"
        f.ls()
```

List of files :
PNA_DATA/Dichro_130627/PNA8xx_Denat_130627-1-200nm-Cell 1.txt
PNA_DATA/Dichro_130627/PNA8xx_Denat_130627-1-200nm-Cell 2.txt
PNA_DATA/Dichro_130627/PNA8xx_Denat_130627-1-200nm-Cell 3.txt
PNA_DATA/Dichro_130627/PNA8xx_Denat_130627-1-217nm-Cell 1.txt
PNA_DATA/Dichro_130627/PNA8xx_Denat_130627-1-217nm-Cell 2.txt
PNA_DATA/Dichro_130627/PNA8xx_Denat_130627-1-217nm-Cell 3.txt
PNA_DATA/Dichro_130627/PNA8xx_Denat_130627-1-230nm-Cell 1.txt
PNA_DATA/Dichro_130627/PNA8xx_Denat_130627-1-230nm-Cell 2.txt
PNA_DATA/Dichro_130627/PNA8xx_Denat_130627-1-230nm-Cell 3.txt
PNA_DATA/Dichro_130627/PNA8xx_Denat_130627-1-260nm-Cell 1.txt

```
PNA_DATA/Dichro_130627/PNA8xx_Denat_130627-1-260nm-Cell 2.txt
PNA_DATA/Dichro_130627/PNA8xx_Denat_130627-1-260nm-Cell 3.txt
PNA_DATA/Dichro_130627/PNA8xx_Denat_130627-1-310nm-Cell 1.txt
PNA_DATA/Dichro_130627/PNA8xx_Denat_130627-1-310nm-Cell 2.txt
PNA_DATA/Dichro_130627/PNA8xx_Denat_130627-1-310nm-Cell 3.txt
PNA_DATA/Dichro_130627/PNA8xx_Denat_130627-1-380nm-Cell 1.txt
PNA_DATA/Dichro_130627/PNA8xx_Denat_130627-1-380nm-Cell 2.txt
PNA_DATA/Dichro_130627/PNA8xx_Denat_130627-1-380nm-Cell 3.txt
```

### 4.1.3 PNA-2 Absorbance

First we fit one wavelength : 200nm

```
In [8]: f.tampon = 1
        f.cell = 2
        f.meth = "ABS"
        f.freq = 200        # the 200nm curve give the best results
        f.analyze()
        Fig("PNA-2 Absorbance")
```

PNA 2 - $\lambda=200\,nm$   Tm: 66.2C +/- 0.94   beta: 0.06   chi2:0.014
Figure S17 : PNA-2 Absorbance

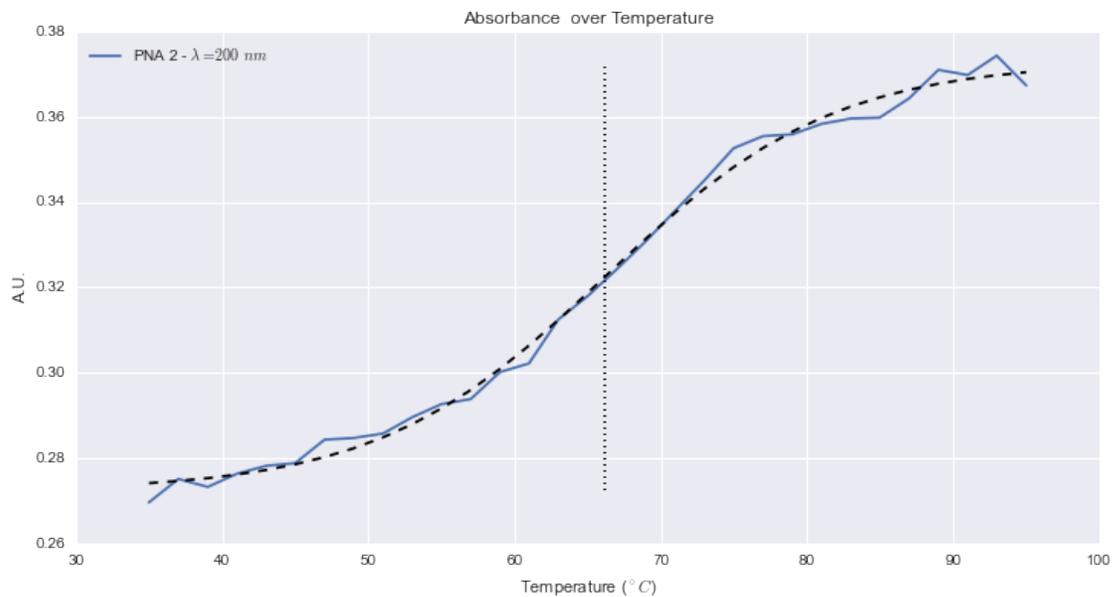

Then we try to fit all the curves at once using `Multiexp()`, defined in multifreq

```
In [9]: f = Multiexp()
        f.BASE = "PNA_DATA/Dichro_130627"
        f.fnom = "PNA8xx_Denat_130627-1-"
        f.tampon = 1
        f.meth = "ABS"
        f.multifreq = (200, 217, 230, 260,310,380)
        f.analyze()
        Fig("PNA-2 Absorbance")
```

```
PNA 2 curve could not be fitted
PNA 2  Tm: 66.3C +/- 1.30    beta: 0.06   chi2:0.048
Figure S18 : PNA-2 Absorbance
```

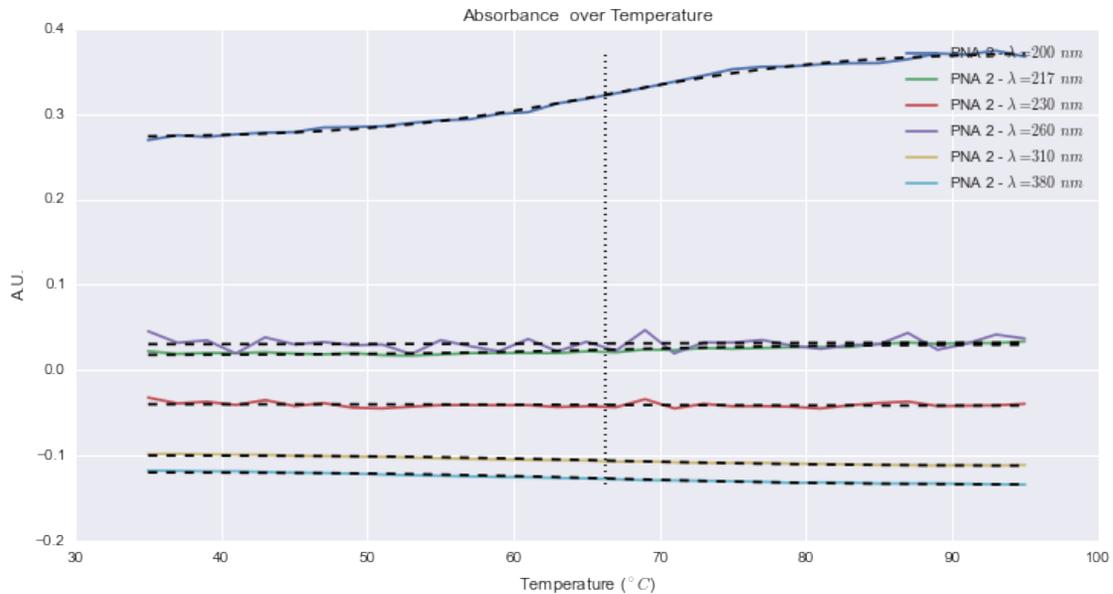

As `Multiexp()` gives satisfactory results, we will use it in the following, all curves are plotted, but only the curve giving interpretable results are fitted.

Results are accumulated into a global ordered dictionary, for the final conclusion

```
In [10]: import collections
         RESULTS = collections.OrderedDict()
         RESULTS["PNA-2 Abs"]=f.Results
```

### 4.1.4  PNA-3 Absorbance

```
In [11]: f.cell = 3
         f.multifreq = (200, 217, 230, 260,310,380)
         f.analyze()
         RESULTS["PNA-3 Abs"] = f.Results
         Fig("PNA-3 Absorbance")
```

```
PNA 2 curve could not be fitted
PNA 2  Tm: 71.1C +/- 1.77    beta: 0.04   chi2:0.054
Figure S19 : PNA-3 Absorbance
```

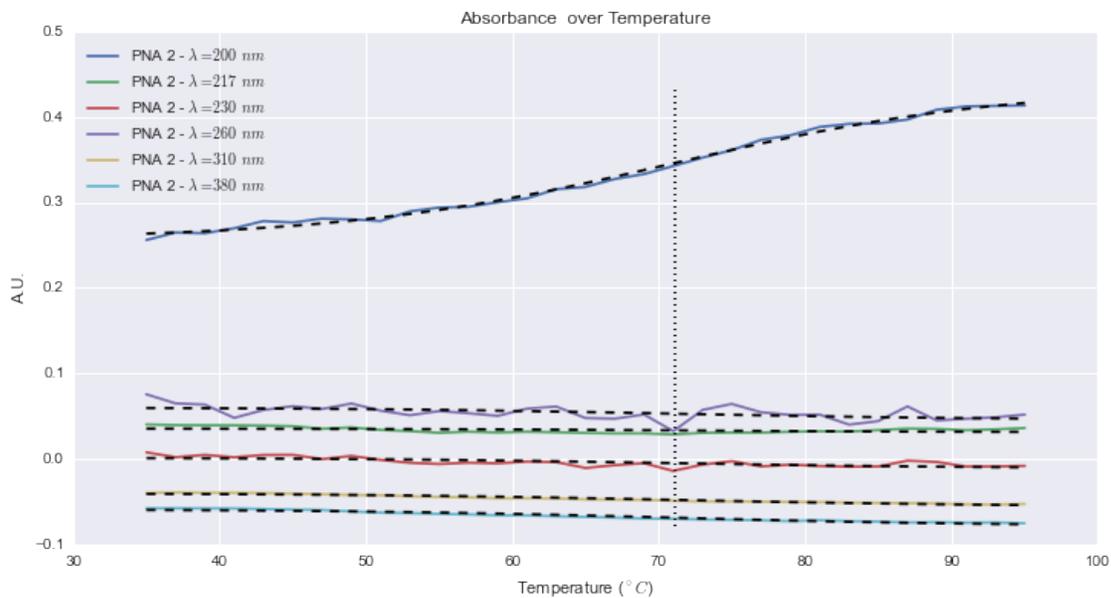

### 4.1.5 PNA-3 CD

```
In [12]: f.meth = "CD"
         f.cell = 3
         f.multifreq = (200, 217, 230, 260, 310, 380)
         f.analyze()
         RESULTS["PNA-3 CD"]=f.Results
         Fig("PNA-3 CD")
```

PNA 2 curve could not be fitted
PNA 2  Tm: 68.4C +/- 0.98   beta: 0.06   chi2:1.572
Figure S20 : PNA-3 CD

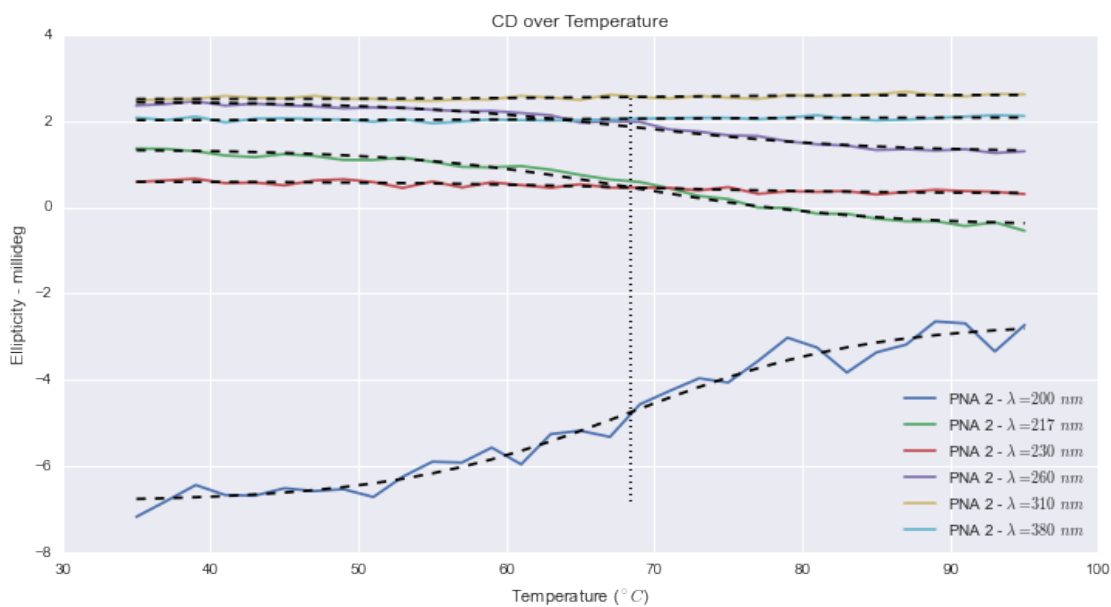

### 4.1.6 Temperature variation, going down

(1 degree / minute)

following wave-lengthes were recorded : 217nm 260nm 380nm

In this series, only CD of PNA-3 could be reliably analyzed, in particular because of the lack of measurements at 200nm, wave-length that gave the best results in the previous series.

Nevertheless, some shift in temperature are observed compared to the previous experiment.

```
In [13]: f = Multiexp()
         f.BASE = "PNA_DATA/Dichro_130627"
         f.fnom = "PNA8xx_Cooling_130627-1-"
         f.ls()
```

```
List of files :
PNA_DATA/Dichro_130627/PNA8xx_Cooling_130627-1-217nm-Cell 1.txt
PNA_DATA/Dichro_130627/PNA8xx_Cooling_130627-1-217nm-Cell 2.txt
PNA_DATA/Dichro_130627/PNA8xx_Cooling_130627-1-217nm-Cell 3.txt
PNA_DATA/Dichro_130627/PNA8xx_Cooling_130627-1-260nm-Cell 1.txt
PNA_DATA/Dichro_130627/PNA8xx_Cooling_130627-1-260nm-Cell 2.txt
PNA_DATA/Dichro_130627/PNA8xx_Cooling_130627-1-260nm-Cell 3.txt
PNA_DATA/Dichro_130627/PNA8xx_Cooling_130627-1-380nm-Cell 1.txt
PNA_DATA/Dichro_130627/PNA8xx_Cooling_130627-1-380nm-Cell 2.txt
PNA_DATA/Dichro_130627/PNA8xx_Cooling_130627-1-380nm-Cell 3.txt
```

### 4.1.7 PNA-2 Absorbance

```
In [14]: f.meth = "ABS"
         f.tampon = 1
         f.cell = 2
         f.multifreq = (217, 260, 380,)
         f.analyze()
         Fig("PNA-2 Absorbance")
```

```
PNA 2 curve could not be fitted
PNA 2  Tm: 54.3C +/- 36.58   beta: 0.03   chi2:0.030
Figure S21 : PNA-2 Absorbance
```

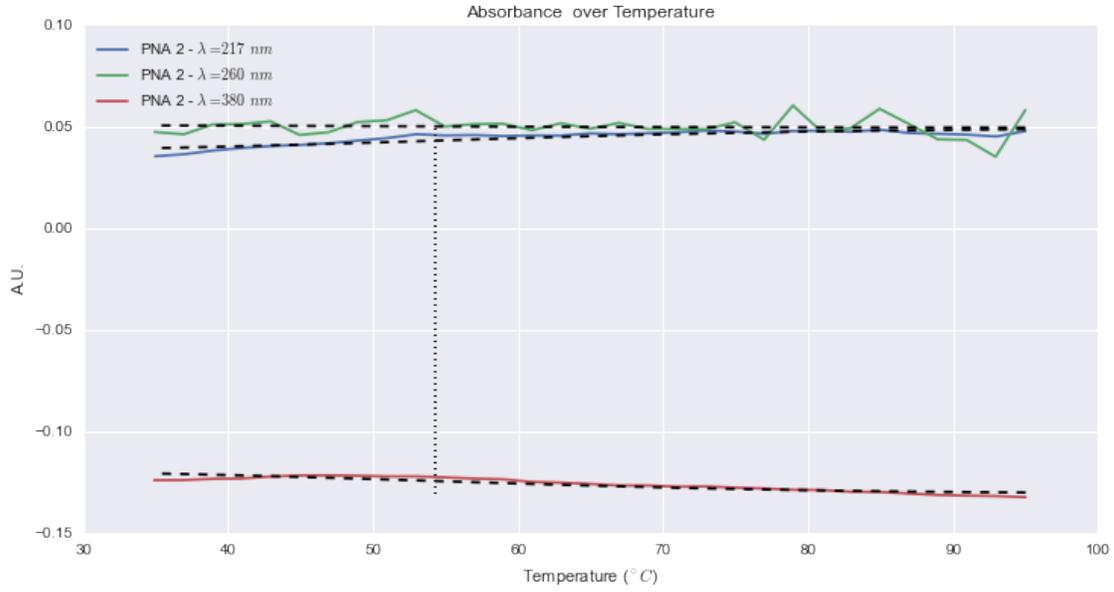

### 4.1.8 PNA-3 Absorbance

```
In [15]: f.meth = "ABS"
         f.tampon = 1
         f.cell = 3
         f.multifreq = (217, 260, 380,)
         f.analyze()
         Fig("PNA-3 Absorbance")
```

PNA 2 curve could not be fitted
PNA 2  Tm: 73.2C +/- 5.00    beta: 0.11   chi2:0.025
Figure S22 : PNA-3 Absorbance

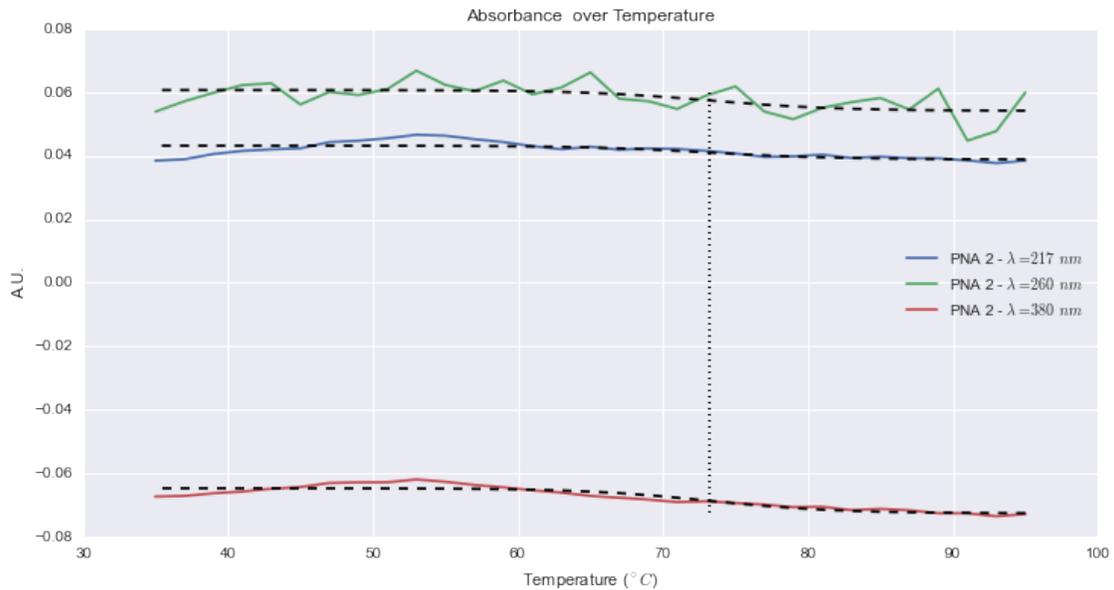

17

### 4.1.9 PNA-3 CD

```
In [16]: f.meth = "CD"
         f.tampon = 1
         f.cell = 3
         f.multifreq = (217, 260, 380,)
         f.analyze()
         RESULTS["PNA-3 CD 2"] = f.Results   # results stored as 2 are for decreasing temperatures
         Fig("PNA-3 CD")
```

PNA 2 curve could not be fitted
PNA 2  Tm: 68.1C +/- 1.82    beta: 0.06   chi2:0.788
Figure S23 : PNA-3 CD

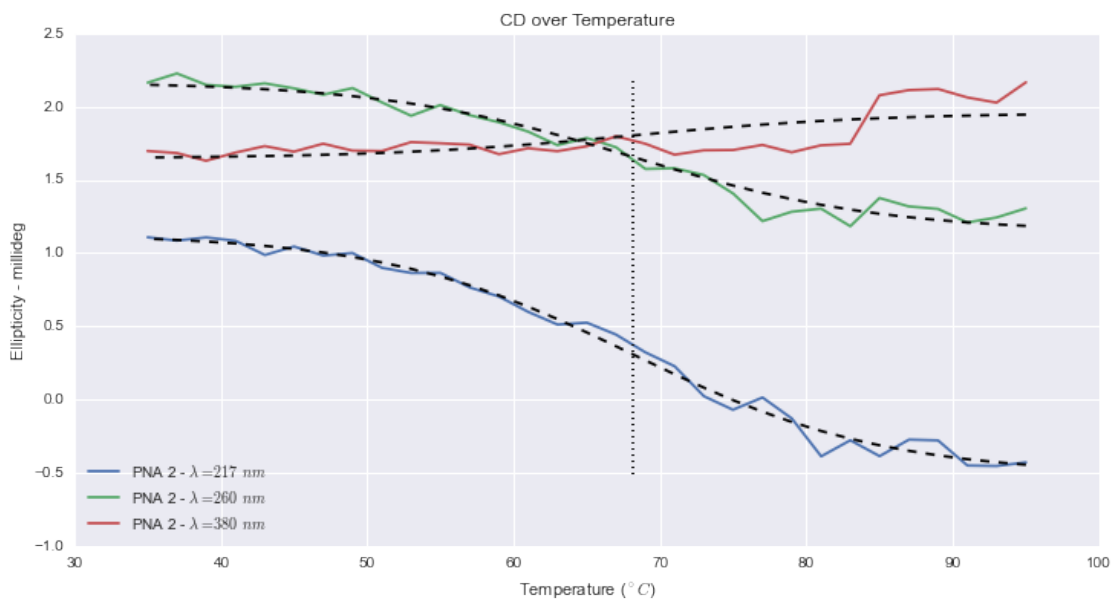

## 4.2 PNA 4 & 6

The same protocole was then applied on the two PNA 4 and 6. Here PNA-4 is in cell 3, and PNA-6 is in cell 4.

PNA-7 (see below) is in cell 5.

### 4.2.1 going up

raison the temperature, 1°C / min
    measured wavelengths 218 240 262 280 350

```
In [17]: f = Multiexp()
         f.BASE = "PNA_DATA/Dichro_130711"
         f.fnom = "PNA8xx_Denat_130711-1-"
         f.ls()
```

```
List of files :
PNA_DATA/Dichro_130711/PNA8xx_Denat_130711-1-218nm-Cell 1.txt
PNA_DATA/Dichro_130711/PNA8xx_Denat_130711-1-218nm-Cell 2.txt
PNA_DATA/Dichro_130711/PNA8xx_Denat_130711-1-218nm-Cell 3.txt
PNA_DATA/Dichro_130711/PNA8xx_Denat_130711-1-218nm-Cell 4.txt
PNA_DATA/Dichro_130711/PNA8xx_Denat_130711-1-218nm-Cell 5.txt
PNA_DATA/Dichro_130711/PNA8xx_Denat_130711-1-240nm-Cell 1.txt
PNA_DATA/Dichro_130711/PNA8xx_Denat_130711-1-240nm-Cell 2.txt
PNA_DATA/Dichro_130711/PNA8xx_Denat_130711-1-240nm-Cell 3.txt
PNA_DATA/Dichro_130711/PNA8xx_Denat_130711-1-240nm-Cell 4.txt
PNA_DATA/Dichro_130711/PNA8xx_Denat_130711-1-240nm-Cell 5.txt
PNA_DATA/Dichro_130711/PNA8xx_Denat_130711-1-262nm-Cell 1.txt
PNA_DATA/Dichro_130711/PNA8xx_Denat_130711-1-262nm-Cell 2.txt
PNA_DATA/Dichro_130711/PNA8xx_Denat_130711-1-262nm-Cell 3.txt
PNA_DATA/Dichro_130711/PNA8xx_Denat_130711-1-262nm-Cell 4.txt
PNA_DATA/Dichro_130711/PNA8xx_Denat_130711-1-262nm-Cell 5.txt
PNA_DATA/Dichro_130711/PNA8xx_Denat_130711-1-280nm-Cell 1.txt
PNA_DATA/Dichro_130711/PNA8xx_Denat_130711-1-280nm-Cell 2.txt
PNA_DATA/Dichro_130711/PNA8xx_Denat_130711-1-280nm-Cell 3.txt
PNA_DATA/Dichro_130711/PNA8xx_Denat_130711-1-280nm-Cell 4.txt
PNA_DATA/Dichro_130711/PNA8xx_Denat_130711-1-280nm-Cell 5.txt
PNA_DATA/Dichro_130711/PNA8xx_Denat_130711-1-350nm-Cell 1.txt
PNA_DATA/Dichro_130711/PNA8xx_Denat_130711-1-350nm-Cell 2.txt
PNA_DATA/Dichro_130711/PNA8xx_Denat_130711-1-350nm-Cell 3.txt
PNA_DATA/Dichro_130711/PNA8xx_Denat_130711-1-350nm-Cell 4.txt
PNA_DATA/Dichro_130711/PNA8xx_Denat_130711-1-350nm-Cell 5.txt
```

### 4.2.2 PNA-4 Absorbance

```
In [18]: f.meth = "ABS"
         f.tampon = 1
         f.cell = 3
         f.nom = 4
         f.multifreq = (218, 280, 240, 262, 350)
         f.analyze()
         RESULTS["PNA-4 Abs"] = f.Results
         Fig("PNA-4 Absorbance")
```

```
PNA 4 curve could not be fitted
PNA 4  Tm: 86.4C +/- 4.18   beta: 0.11  chi2:0.059
Figure S24 : PNA-4 Absorbance
```

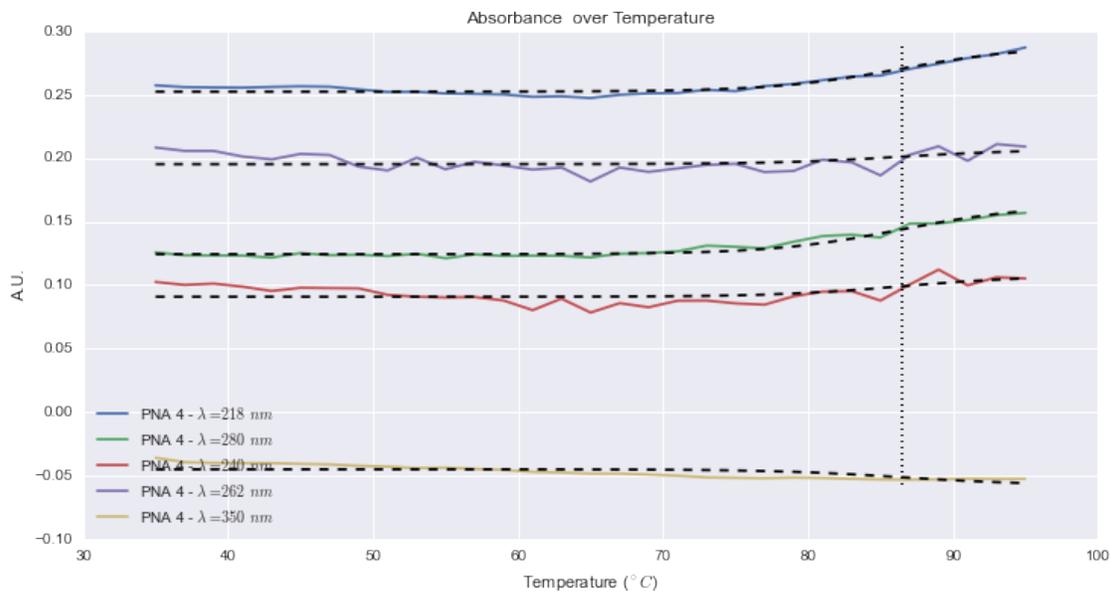

### 4.2.3 PNA-6 Absorbance

```
In [19]: f.cell = 4
         f.nom = 6
         f.multifreq = (218, 280, 240, 262, 350)
         f.analyze()
         RESULTS["PNA-6 Abs"] = f.Results
         Fig("PNA-6 Absorbance")
```

```
PNA 6 curve could not be fitted
PNA 6  Tm: 88.8C +/- 5.51    beta: 0.09   chi2:0.060
Figure S25 : PNA-6 Absorbance
```

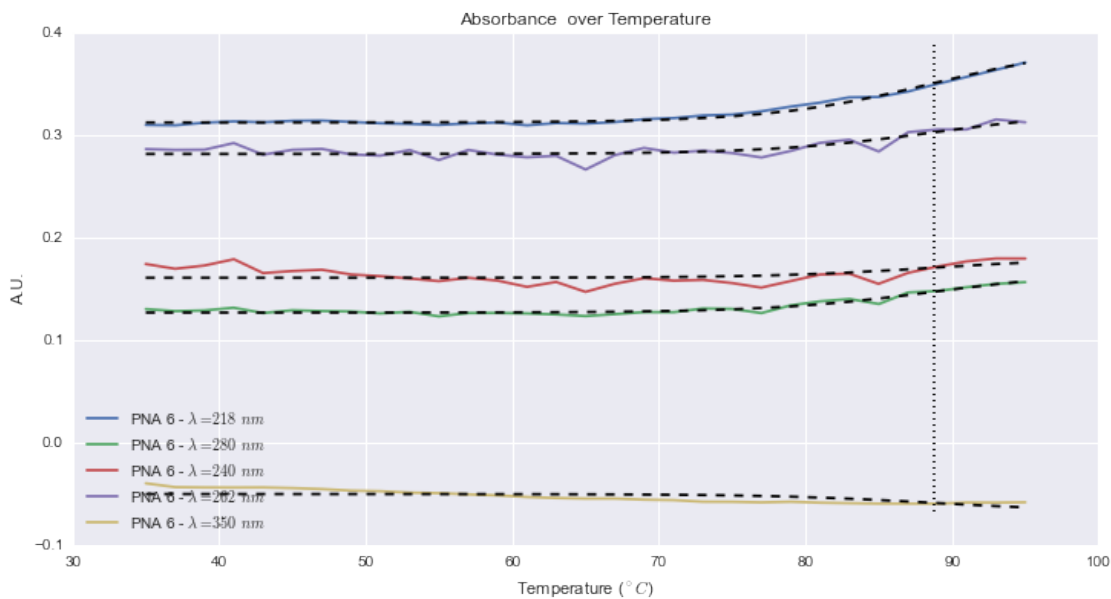

### 4.2.4 PNA-4 by CD

```
In [20]: f.meth = "CD"
         f.cell = 3
         f.nom = 4
         f.multifreq = (218, 280, 240, 262, 350)
         f.analyze()
         RESULTS["PNA-4 CD"] = f.Results
         Fig("PNA-4 CD")
```

```
PNA 4 curve could not be fitted
PNA 4  Tm: 81.7C +/- 2.43    beta: 0.04   chi2:0.750
Figure S26 : PNA-4 CD
```

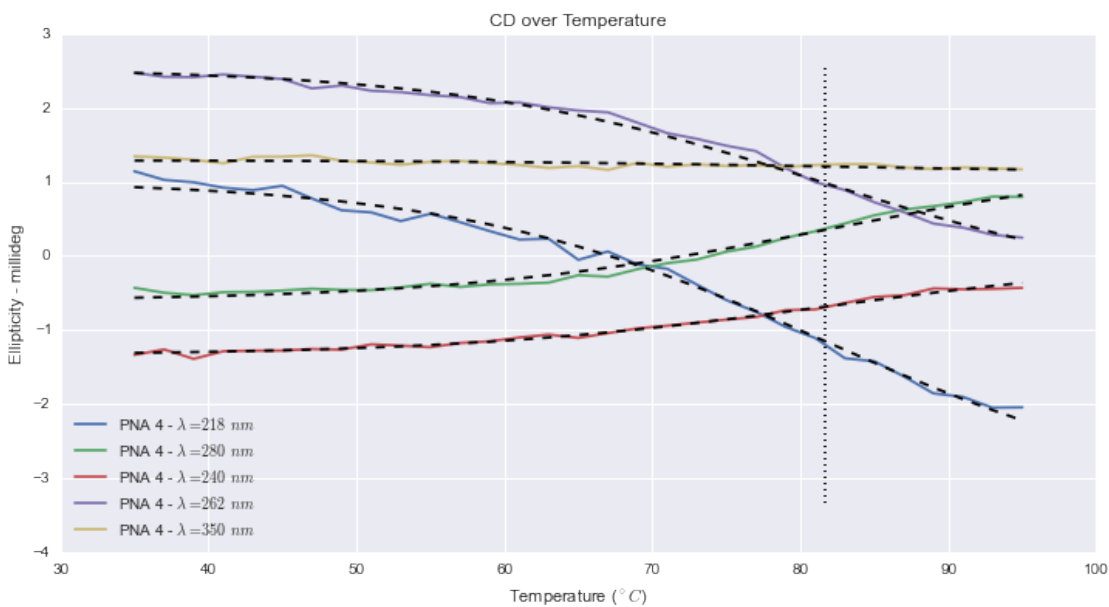

### 4.2.5 PNA-6 by CD

```
In [21]: f.cell = 4
         f.nom = 6
         f.multifreq = (218, 280, 240, 262, 350)
         f.analyze()
         RESULTS["PNA-6 CD"] = f.Results
         Fig("PNA-6 CD")
```

```
PNA 6 curve could not be fitted
PNA 6  Tm: 83.0C +/- 4.50    beta: 0.03   chi2:1.108
Figure S27 : PNA-6 CD
```

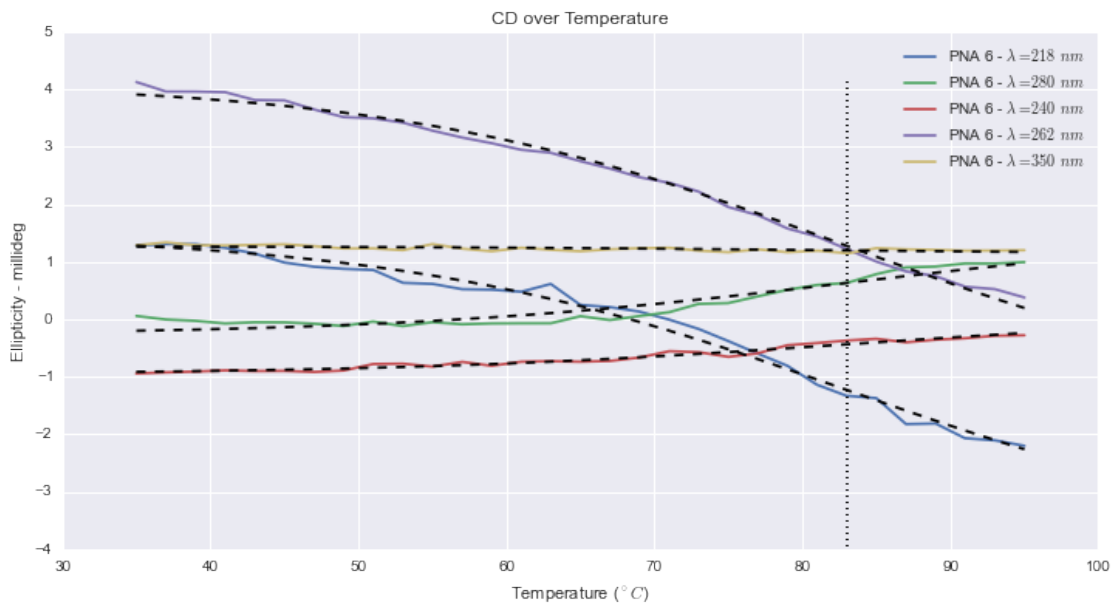

### 4.2.6 going down

218 240 262 280 350 Abs does not work

```
In [22]: f = Multiexp()
         f.BASE = "PNA_DATA/Dichro_130711"
         f.fnom = "PNA8xx_Cooling_130711-1-"
         f.ls()
```

```
List of files :
PNA_DATA/Dichro_130711/PNA8xx_Cooling_130711-1-218nm-Cell 1.txt
PNA_DATA/Dichro_130711/PNA8xx_Cooling_130711-1-218nm-Cell 2.txt
PNA_DATA/Dichro_130711/PNA8xx_Cooling_130711-1-218nm-Cell 3.txt
PNA_DATA/Dichro_130711/PNA8xx_Cooling_130711-1-218nm-Cell 4.txt
PNA_DATA/Dichro_130711/PNA8xx_Cooling_130711-1-218nm-Cell 5.txt
PNA_DATA/Dichro_130711/PNA8xx_Cooling_130711-1-240nm-Cell 1.txt
PNA_DATA/Dichro_130711/PNA8xx_Cooling_130711-1-240nm-Cell 2.txt
PNA_DATA/Dichro_130711/PNA8xx_Cooling_130711-1-240nm-Cell 3.txt
PNA_DATA/Dichro_130711/PNA8xx_Cooling_130711-1-240nm-Cell 4.txt
PNA_DATA/Dichro_130711/PNA8xx_Cooling_130711-1-240nm-Cell 5.txt
PNA_DATA/Dichro_130711/PNA8xx_Cooling_130711-1-262nm-Cell 1.txt
PNA_DATA/Dichro_130711/PNA8xx_Cooling_130711-1-262nm-Cell 2.txt
PNA_DATA/Dichro_130711/PNA8xx_Cooling_130711-1-262nm-Cell 3.txt
PNA_DATA/Dichro_130711/PNA8xx_Cooling_130711-1-262nm-Cell 4.txt
PNA_DATA/Dichro_130711/PNA8xx_Cooling_130711-1-262nm-Cell 5.txt
PNA_DATA/Dichro_130711/PNA8xx_Cooling_130711-1-280nm-Cell 1.txt
PNA_DATA/Dichro_130711/PNA8xx_Cooling_130711-1-280nm-Cell 2.txt
PNA_DATA/Dichro_130711/PNA8xx_Cooling_130711-1-280nm-Cell 3.txt
PNA_DATA/Dichro_130711/PNA8xx_Cooling_130711-1-280nm-Cell 4.txt
PNA_DATA/Dichro_130711/PNA8xx_Cooling_130711-1-280nm-Cell 5.txt
PNA_DATA/Dichro_130711/PNA8xx_Cooling_130711-1-350nm-Cell 1.txt
PNA_DATA/Dichro_130711/PNA8xx_Cooling_130711-1-350nm-Cell 2.txt
```

```
PNA_DATA/Dichro_130711/PNA8xx_Cooling_130711-1-350nm-Cell 3.txt
PNA_DATA/Dichro_130711/PNA8xx_Cooling_130711-1-350nm-Cell 4.txt
PNA_DATA/Dichro_130711/PNA8xx_Cooling_130711-1-350nm-Cell 5.txt

In [23]: f.meth = "CD"
         f.tampon = 1
         f.cell = 3
         f.nom = 4
         f.multifreq = (218, 240, 262, 280, 350)
         f.analyze()
         RESULTS["PNA-4 CD 2"] = f.Results
         Fig("PNA-4 CD")
```

PNA 4 curve could not be fitted
PNA 4  Tm: 80.2C +/- 1.70    beta: 0.04   chi2:0.731
Figure S28 : PNA-4 CD

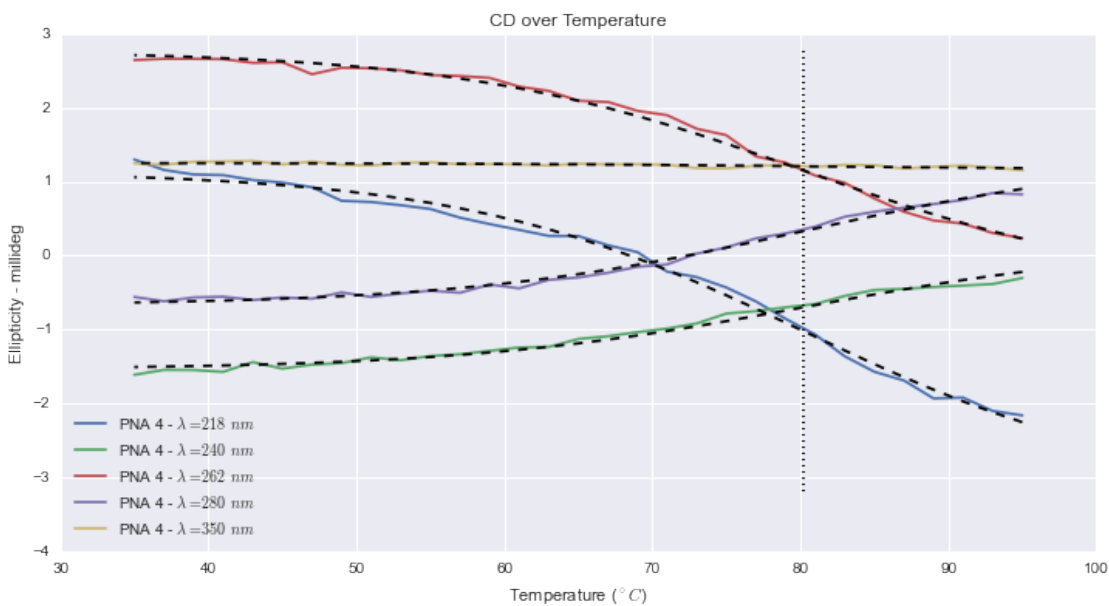

```
In [24]: f.meth = "CD"
         f.tampon = 1
         f.cell = 4
         f.nom = 6
         f.multifreq = (218, 240, 262, 280, 350)
         f.analyze()
         RESULTS["PNA-6 CD 2"] = f.Results
         Fig("PNA-6 CD")
```

PNA 6 curve could not be fitted
PNA 6  Tm: 80.9C +/- 3.60    beta: 0.03   chi2:0.996
Figure S29 : PNA-6 CD

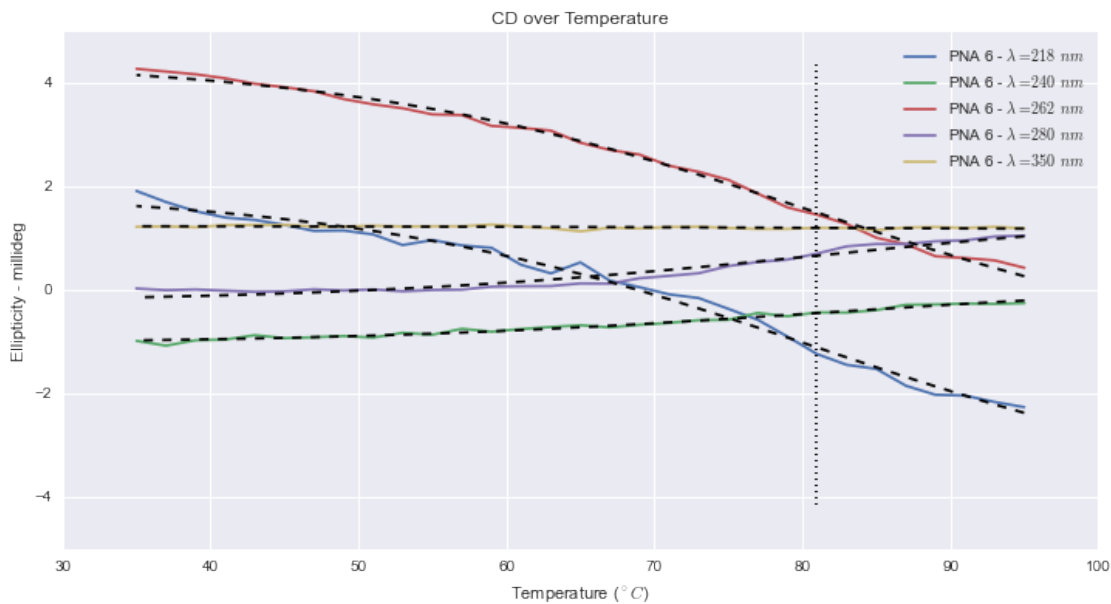

## 4.3 PNA 5

Buffer is in cell 1 and PNA-5 in cell 3
    PNA-5 is non-chiral, so CD is irrelevant
    wavelengths : 200 207 215 250

```
In [25]: f = Multiexp()
         f.BASE = "PNA_DATA/Dichro_PNA5/Abs_Te/ThermalDenat_120919/"
         f.fnom = "PNA_120919_ThermDenat-1-"
         f.extension = ".csv"
         f.ls()

List of files :
PNA_DATA/Dichro_PNA5/Abs_Te/ThermalDenat_120919/PNA_120919_ThermDenat-1-200nm-cell 1.csv
PNA_DATA/Dichro_PNA5/Abs_Te/ThermalDenat_120919/PNA_120919_ThermDenat-1-200nm-cell 3.csv
PNA_DATA/Dichro_PNA5/Abs_Te/ThermalDenat_120919/PNA_120919_ThermDenat-1-207nm-cell 1.csv
PNA_DATA/Dichro_PNA5/Abs_Te/ThermalDenat_120919/PNA_120919_ThermDenat-1-207nm-cell 3.csv
PNA_DATA/Dichro_PNA5/Abs_Te/ThermalDenat_120919/PNA_120919_ThermDenat-1-215nm-cell 1.csv
PNA_DATA/Dichro_PNA5/Abs_Te/ThermalDenat_120919/PNA_120919_ThermDenat-1-215nm-cell 3.csv
PNA_DATA/Dichro_PNA5/Abs_Te/ThermalDenat_120919/PNA_120919_ThermDenat-1-250nm-cell 1.csv
PNA_DATA/Dichro_PNA5/Abs_Te/ThermalDenat_120919/PNA_120919_ThermDenat-1-250nm-cell 3.csv

In [26]: f.meth = "ABS"
         f.tampon = 1
         f.cell = 3
         f.nom = 5
         f.multifreq = (200, 207, 215, 250)
         f.analyze()
         RESULTS["PNA-5 Abs"]=f.Results
         Fig("PNA-5 Absorbance")
```

```
PNA 5 curve could not be fitted
PNA 5  Tm: 73.5C +/- 0.54    beta: 0.04   chi2:0.148
Figure S30 : PNA-5 Absorbance
```

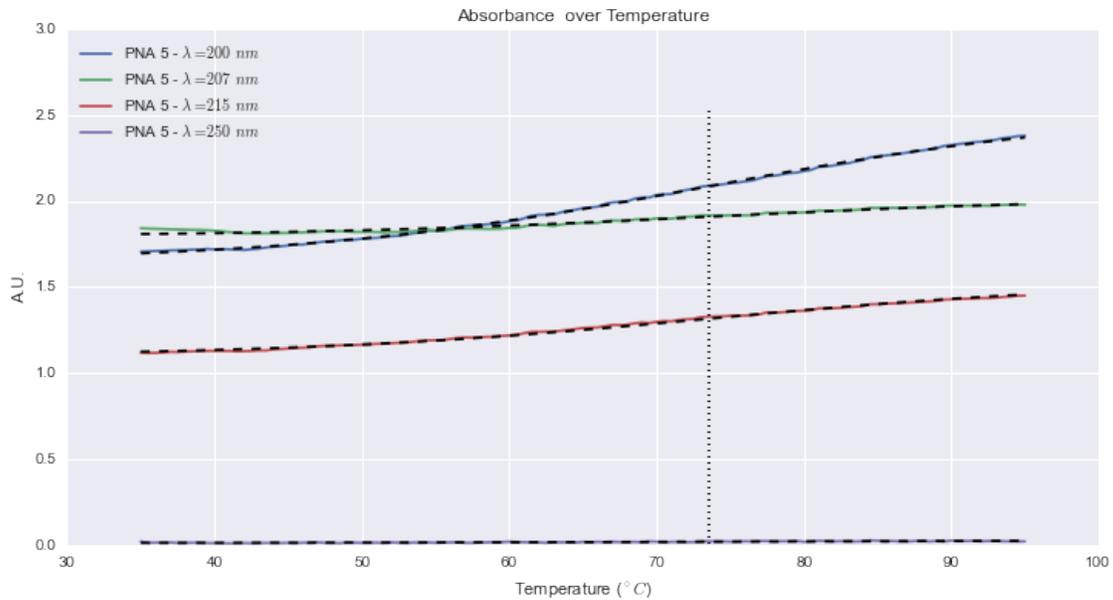

```
In [27]: f = Multiexp()
         f.BASE = "PNA_DATA/Dichro_PNA5/Abs_Te/ThermalCooling_120920/"
         f.fnom = "PNA_120920_ThermCool-1-"
         f.extension = ".csv"
         #f.ls()
         f.meth = "ABS"
         f.tampon = 1
         f.cell = 3
         f.nom = 5
         f.multifreq = (200, 207, 215, 250)
         f.analyze()
         RESULTS["PNA-5 Abs 2"]=f.Results
         Fig("PNA-5 Absorbance")

PNA 5 curve could not be fitted
PNA 5  Tm: 70.8C +/- 1.77    beta: 0.03   chi2:0.317
Figure S31 : PNA-5 Absorbance
```

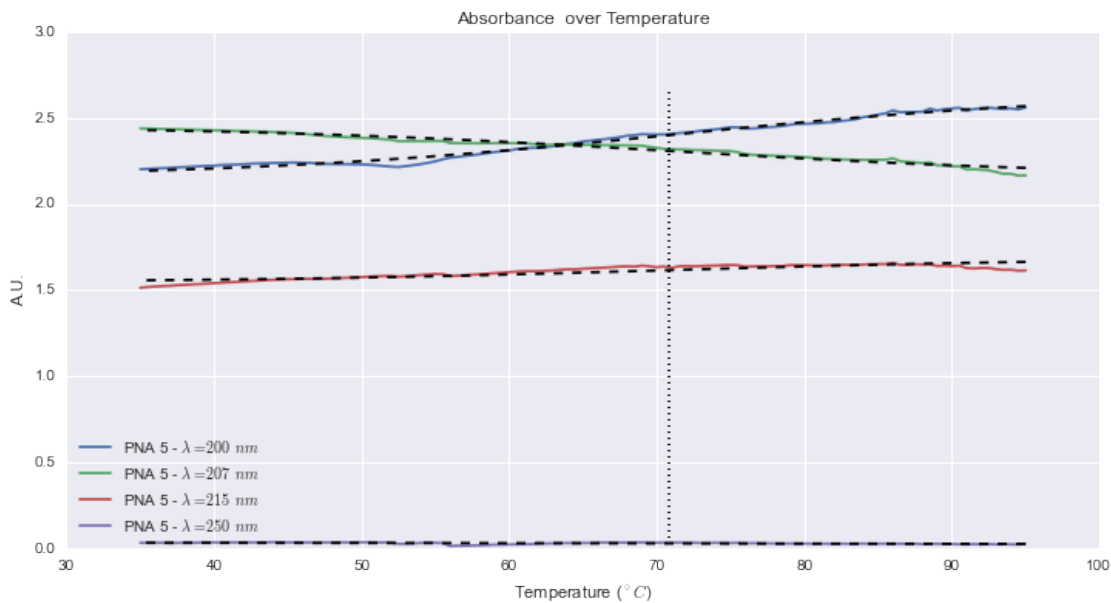

## 4.4 PNA 7

PNA 7 is quite special, it presents a $T_m$ very high, much over than the highest experimental temperature tested. In consequence, the experimental curves only sample a small portion of the melting curve, and do not permit a precise evaluation of the melting temperature.

It can be seen in the curves, and the results given for the different experiment are not coherent.

```
In [28]: f = Multiexp()
         f.BASE = "PNA_DATA/Dichro_130711"
         f.fnom = "PNA8xx_Denat_130711-1-"
         f.ls()
         f.meth = "ABS"
         f.tampon = 1
         f.cell = 5
         f.nom = 7
         f.multifreq = (218, 240, 262, 280) #, 350)
         f.analyze()
         RESULTS["PNA-7 Abs"]=f.Results
         Fig("PNA-7 Absorbance")
```

```
List of files :
PNA_DATA/Dichro_130711/PNA8xx_Denat_130711-1-218nm-Cell 1.txt
PNA_DATA/Dichro_130711/PNA8xx_Denat_130711-1-218nm-Cell 2.txt
PNA_DATA/Dichro_130711/PNA8xx_Denat_130711-1-218nm-Cell 3.txt
PNA_DATA/Dichro_130711/PNA8xx_Denat_130711-1-218nm-Cell 4.txt
PNA_DATA/Dichro_130711/PNA8xx_Denat_130711-1-218nm-Cell 5.txt
PNA_DATA/Dichro_130711/PNA8xx_Denat_130711-1-240nm-Cell 1.txt
PNA_DATA/Dichro_130711/PNA8xx_Denat_130711-1-240nm-Cell 2.txt
PNA_DATA/Dichro_130711/PNA8xx_Denat_130711-1-240nm-Cell 3.txt
PNA_DATA/Dichro_130711/PNA8xx_Denat_130711-1-240nm-Cell 4.txt
PNA_DATA/Dichro_130711/PNA8xx_Denat_130711-1-240nm-Cell 5.txt
```

```
PNA_DATA/Dichro_130711/PNA8xx_Denat_130711-1-262nm-Cell 1.txt
PNA_DATA/Dichro_130711/PNA8xx_Denat_130711-1-262nm-Cell 2.txt
PNA_DATA/Dichro_130711/PNA8xx_Denat_130711-1-262nm-Cell 3.txt
PNA_DATA/Dichro_130711/PNA8xx_Denat_130711-1-262nm-Cell 4.txt
PNA_DATA/Dichro_130711/PNA8xx_Denat_130711-1-262nm-Cell 5.txt
PNA_DATA/Dichro_130711/PNA8xx_Denat_130711-1-280nm-Cell 1.txt
PNA_DATA/Dichro_130711/PNA8xx_Denat_130711-1-280nm-Cell 2.txt
PNA_DATA/Dichro_130711/PNA8xx_Denat_130711-1-280nm-Cell 3.txt
PNA_DATA/Dichro_130711/PNA8xx_Denat_130711-1-280nm-Cell 4.txt
PNA_DATA/Dichro_130711/PNA8xx_Denat_130711-1-280nm-Cell 5.txt
PNA_DATA/Dichro_130711/PNA8xx_Denat_130711-1-350nm-Cell 1.txt
PNA_DATA/Dichro_130711/PNA8xx_Denat_130711-1-350nm-Cell 2.txt
PNA_DATA/Dichro_130711/PNA8xx_Denat_130711-1-350nm-Cell 3.txt
PNA_DATA/Dichro_130711/PNA8xx_Denat_130711-1-350nm-Cell 4.txt
PNA_DATA/Dichro_130711/PNA8xx_Denat_130711-1-350nm-Cell 5.txt
PNA 7 curve could not be fitted
PNA 7  Tm: 90.2C +/- 5.25   beta: 0.16   chi2:0.053
Figure S32 : PNA-7 Absorbance
```

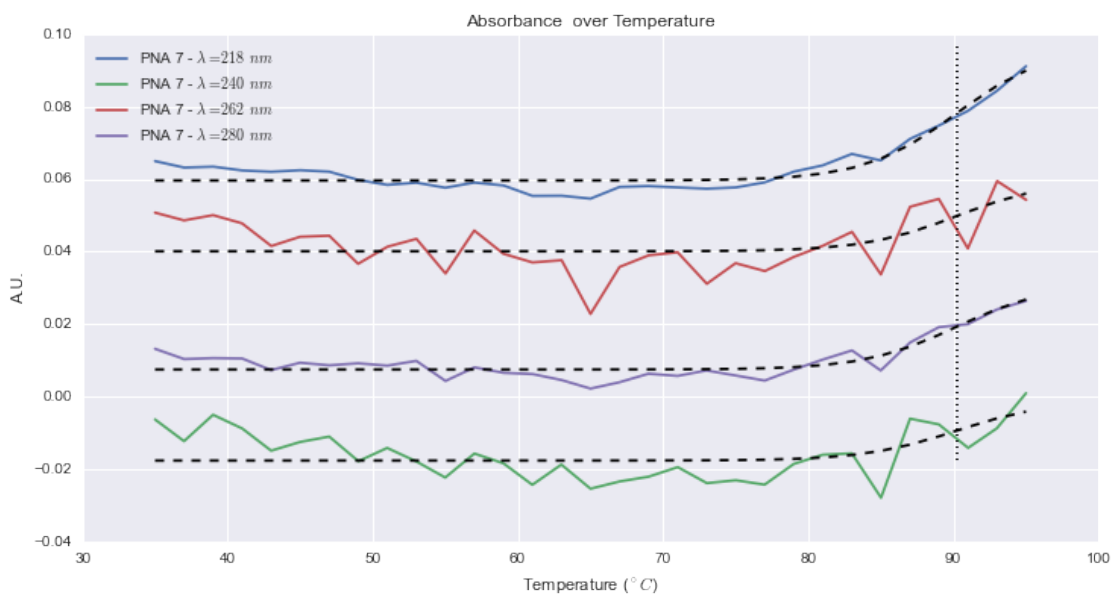

```
In [29]: f.meth = "CD"
         f.multifreq = (218, 240, 262, 280)
         f.analyze(Tm=80.0,beta=0.05)
         Fig("PNA-7 CD")
```

Figure S33 : PNA-7 CD

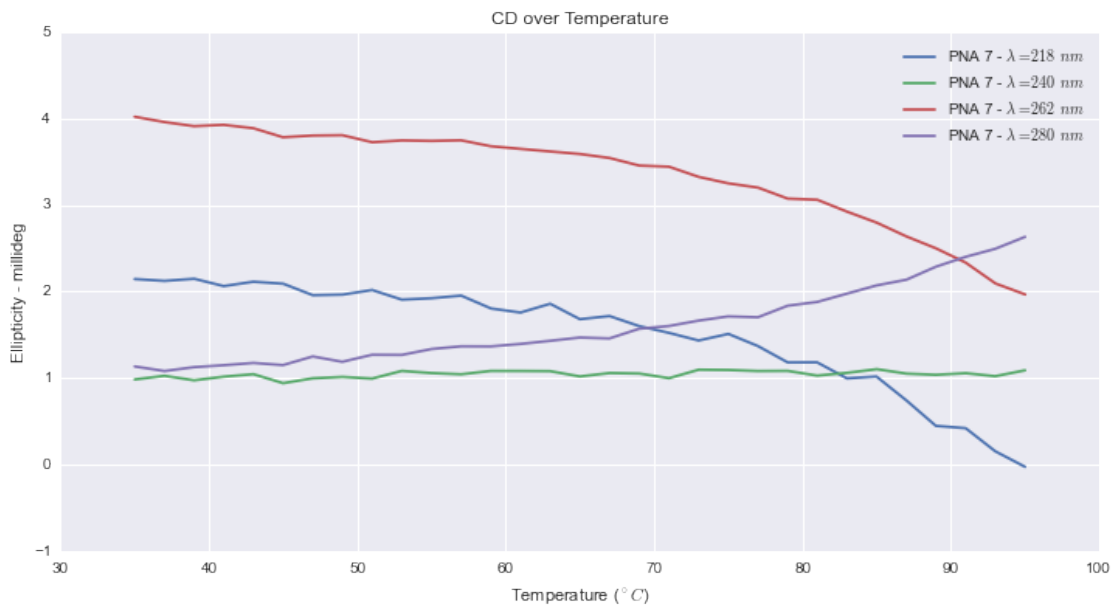

Here, the program is not able to fit the data, even though the initial values (plotted above) seem pretty good.

Probably the $T_m$ value is very high, and cannot be determined.

```
In [30]: f = Multiexp()
         f.BASE = "PNA_DATA/Dichro_130711"
         f.fnom = "PNA8xx_Cooling_130711-1-"
         #f.ls()
         f.meth = "ABS"
         f.tampon = 1
         f.cell = 5
         f.nom = 7
         f.multifreq = (218,  240, 262, 280)
         f.analyze()
         RESULTS["PNA-7 Abs 2"]=f.Results
         Fig("PNA-7 Absorbance")
```

```
PNA 7 curve could not be fitted
PNA 7   Tm: 72.1C +/- 3.24    beta: 0.15   chi2:0.064
Figure S34 : PNA-7 Absorbance
```

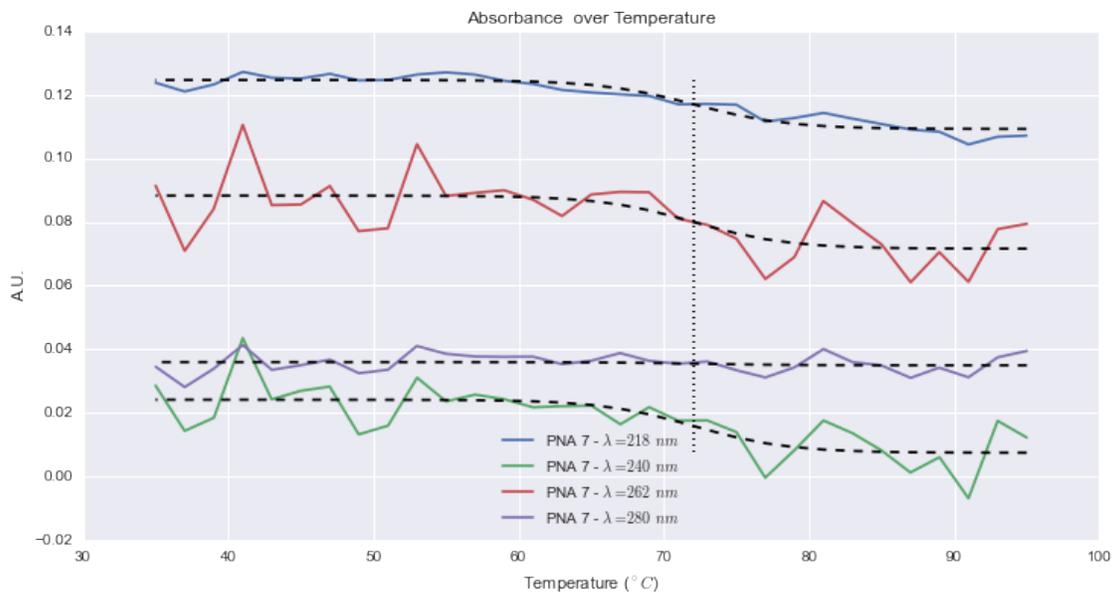

```
In [31]: f.meth = "CD"
         f.multifreq = (218, 240, 262, 280, 350)
         f.analyze()
         RESULTS["PNA-7 CD 2"]=f.Results
         Fig("PNA-7 CD")
```

PNA 7 curve could not be fitted
PNA 7  Tm: 165.5C +/- 149.85    beta: 0.02   chi2:0.501
Figure S35 : PNA-7 CD

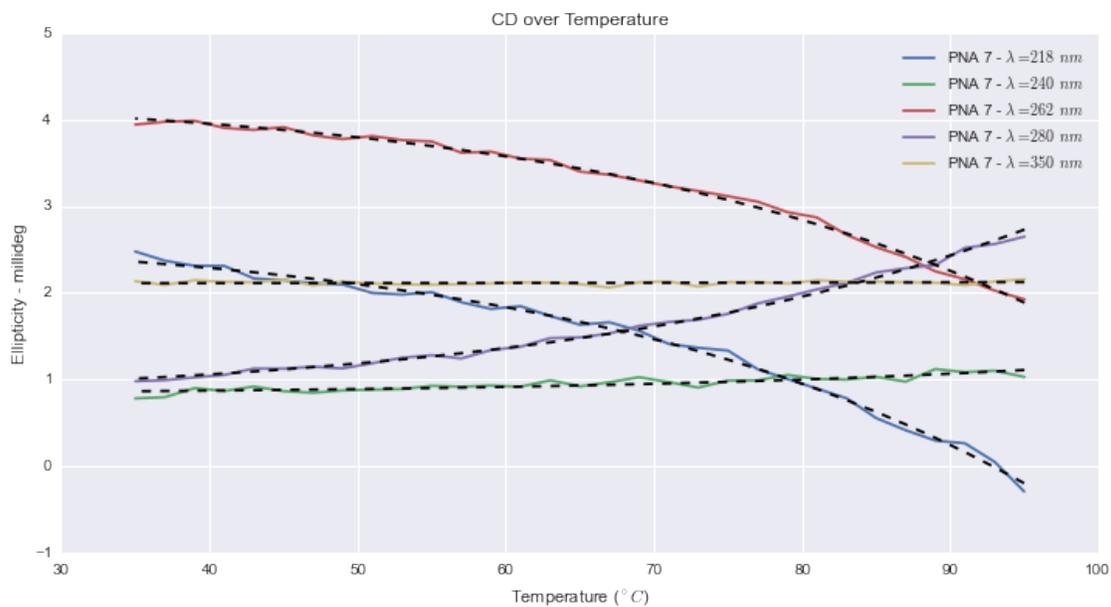

29

# 5 conclusion

Results have been accumulated and are given below:

```
In [32]: #print RESULTS
         for k,v in RESULTS.items():
             print "%s Tm : %.1f +/- %.2f"%(k, v['Tm'], v['Tm_errorbar'])

PNA-2 Abs Tm : 66.3 +/- 1.30
PNA-3 Abs Tm : 71.1 +/- 1.77
PNA-3 CD Tm : 68.4 +/- 0.98
PNA-3 CD 2 Tm : 68.1 +/- 1.82
PNA-4 Abs Tm : 86.4 +/- 4.18
PNA-6 Abs Tm : 88.8 +/- 5.51
PNA-4 CD Tm : 81.7 +/- 2.43
PNA-6 CD Tm : 83.0 +/- 4.50
PNA-4 CD 2 Tm : 80.2 +/- 1.70
PNA-6 CD 2 Tm : 80.9 +/- 3.60
PNA-5 Abs Tm : 73.5 +/- 0.54
PNA-5 Abs 2 Tm : 70.8 +/- 1.77
PNA-7 Abs Tm : 90.2 +/- 5.25
PNA-7 Abs 2 Tm : 72.1 +/- 3.24
PNA-7 CD 2 Tm : 165.5 +/- 149.85
```



\end{document}